\documentclass[tex,twocolumn,epjc3,tightenlines]{svjour3}
\RequirePackage[T1]{fontenc}
\RequirePackage{graphicx}
\RequirePackage{mathptmx}      
\RequirePackage[numbers,sort&compress]{natbib}
\RequirePackage[colorlinks,citecolor=blue,urlcolor=blue,linkcolor=blue]{hyperref}
\usepackage{kantlipsum,widetext}
\usepackage{amsmath, amssymb}
\usepackage{graphics,bm}
\usepackage{graphicx}
\usepackage{bbold}
\usepackage{slashed}
\usepackage{feynmf}
\usepackage{physics}
\usepackage{wrapfig}
\usepackage{hyperref}
\usepackage[toc,page]{appendix}

\usepackage{tikz}
\usetikzlibrary{arrows,shapes}
\usetikzlibrary{trees}
\usetikzlibrary{matrix,arrows} 			
\usetikzlibrary{positioning}				
\usetikzlibrary{calc,through}				
\usepackage{pgffor}							

\usetikzlibrary{decorations.pathmorphing}	
\usetikzlibrary{decorations.markings}
\tikzset{
    sigmaCT/.style={draw=black, postaction={decorate},
        decoration={markings,mark=at position .99 with {\arrow[draw=black]{>}},mark=at 		 position .99 with {\arrow[draw=black]{<}}}},
    pionCT/.style={dashed,draw=black, postaction={decorate},
        decoration={markings,mark=at position .99 with {\arrow[draw=black]{>}},mark=at position .99 with {\arrow[draw=black]{<}}}},    
    fermionCT/.style={draw=black, postaction={decorate},
        decoration={markings,mark=at position .5 with {\arrow[draw=black]{>}},mark=at position .99 with {\arrow[draw=black]{>}},mark=at position .99 with {\arrow[draw=black]{<}}}},    
    fermion/.style={draw=black, postaction={decorate},
        decoration={markings,mark=at position .55 with {\arrow[draw=black]{>}}}},
    fermionbar/.style={draw=black, postaction={decorate},
        decoration={markings,mark=at position .55 with {\arrow[draw=black]{<}}}},
    pion/.style={dashed,draw=black, postaction={decorate}},
    sigma/.style={draw=black, postaction={decorate}}
}

\newcommand{\be}{\begin{equation}}
\newcommand{\ee}{\end{equation}}

\newcommand{\bqa}{\begin{eqnarray}}
\newcommand{\eqa}{\end{eqnarray}}

\newcommand{\os}{\text{\tiny OS}}
\newcommand{\ms}{\overline{\text{\tiny MS}}}

\def\square{\vcenter{\vbox{\hrule height.4pt
          \hbox{\vrule width.4pt height4pt
          \kern4pt\vrule width.3pt}\hrule height.4pt}}}

\journalname{Eur. Phys. J.}

\begin{document}

\title{QCD phase diagram in a constant magnetic background}
\subtitle{Inverse magnetic catalysis: where models meet the lattice }

\author{Jens O. Andersen\thanksref{e2,addr2}}
\thankstext{e2}{e-mail: andersen@tf.phys.ntnu.no} 

\institute{
  Department of Physics, 
Norwegian University of Science and Technology, H{\o}gskoleringen 5,
N-7491 Trondheim, Norway
\label{addr2}
}
\date{Received: date / Revised version: date}

\maketitle

\begin{abstract}
Magnetic catalysis is the enhancement of a condensate 
due to the presence of an external magnetic field.
Magnetic catalysis at $T=0$ is a robust phenomenon in low-energy theories and models of QCD
as well as in lattice simulations. We review the underlying physics of magnetic catalysis from both perspectives. 
The quark-meson model is used as a specific example of a model
that exhibits magnetic catalysis.
Regularization and renormalization are discussed  
and we pay particular attention to a consistent and correct determination of the parameters of the Lagrangian using the on-shell renormalization scheme.
A straightforward application of the quark-meson model and the
NJL model leads to the prediction that the
chiral transition temperature $T_{\chi}$ is increasing as a function of the magnetic field $B$. This is in 
disagreement with lattice results, which show that $T_{\chi}$ is a decreasing function of $B$, independent of the pion mass.
The behavior can be understood in terms of the so-called
valence and sea contributions to the quark condensate
and the competition between them.
We critically examine these ideas as well recent attempts to improve low-energy models using lattice input.

\end{abstract}

\section{Introduction}
\label{intro}
The phase diagram of QCD has received a lot of attention since the first ideas appeared in the 1970s. At that time, it was thought that QCD has two phases, a hadronic phase at low temperatures and a deconfined phase of quarks and gluons at high temperatures. 
In 1984, Bailin and Love~\cite{colorfirst} suggested that at high density,
quark matter should be a color superconductor. The ideas are analogous to those of ordinary superconductivity and BCS theory~\cite{bcs}, namely the instability of the 
Fermi surface to form Cooper pairs under an attractive interaction.
In QCD, an attractive interaction is provided by one-gluon exchange in the triplet channel. Since then, there have been large efforts to map out the phase diagram of QCD and study the properties of its different
phases~\cite{raja,alford,fukurev}. The phase diagram has shown to be surprisingly rich at high baryon density and low temperatures.
It includes a quarkyonic phase~\cite{rob1} 
as well as a number of superconducting phases,
some of them being inhomogeneous. Most of these results have been obtained using low-energy models of QCD, notably the quark-meson (QM) model and the Nambu-Jona-Lasinio (NJL) model, with or without coupling to the Polyakov loop.
The reason is that lattice simulations are notoriously difficult to perform at finite baryon chemical potential $\mu_B$ due to the sign problem, so that one cannot use techniques involving importance sampling.

The temperature $T$ and baryon chemical potential $\mu_B$ are not the only
relevant parameters of QCD. For example, one can introduce a separate chemical potential $\mu_f$ for each quark flavor $f$. For two flavors, this
leads to another independent chemical potential besides $\mu_B$, namely the isospin chemical potential $\mu_I$. For the three flavors, $\mu_S={1\over2}(\mu_u+\mu_d-2\mu_s)$ is added.
The addition of these chemical potentials gives rise to pion and kaon condensation. At $T=0$, pion condensation occurs for $\mu_I>m_{\pi}$, while kaon condensation takes place for $|\pm{1\over2}\mu_I+\mu_S|>m_K$ (upper sign for charged kaons and lower sign for neutral kaons).
The former is particularly interesting since finite $\mu_I$ and vanishing $\mu_B$
has no sign problem and is therefore amenable to lattice simulations.

The final example of an external parameter, which is the topic of this review,
is a (constant) magnetic background. There are several areas of high-energy physics,
where such a background is relevant. One is non-central heavy-ion collisions,
where large, time-dependent fields are generated. These fields are short-lived and have a maximum value of approximately $eB=6m_{\pi}^2$~\cite{kar1}.
The basis mechanism is simply that (in the center-of mass frame)
the two nuclei represent electric currents that according to Maxwell's equations generate a magnetic field.
Another example where strong magnetic fields appear, are 
magnetars~\cite{duncan}. This is a special class of neutron stars
with relatively low rotation frequencies.
It is believed that the magnetic fields on the
surface are $10^{14}-10^{15}$ Gauss, while in the interior they can be as strong
as $10^{16}-10^{19}$ Gauss.

We consider QCD with an $SU(3)$ gauge group, a global $SU(N_f)$
vector symmetry and quark masses $m_f$.
The QCD Lagrangian is
\bqa\nonumber
{\cal L }_{\rm QCD}&=&
-{1\over4}F_{\mu\nu}^aF_a^{\mu\nu}
+i\bar{\psi}_f\gamma^{\mu}D_{\mu}\psi_f
-m_f\bar{\psi}_f\psi_f
\\ &&
+{\cal L}_{\rm gf}+{\cal L}_{\rm ghost}\;,
\label{qcdl}
\eqa
where the gluon field strength tensor is $F^a_{\mu\nu}=\partial_{\mu}A_{\nu}^a-\partial_{\nu}A^a_{\mu}
-gf^{abc}A_{\mu}^bA_{\nu}^c$, $f^{abc}$
are the structure constants
the  covariant derivative in the presence of an abelian background field
$A_{\mu}^{\rm EM}$ is
\bqa
D_{\mu}&=&\partial_{\mu}+iq_fA_{\mu}^{\rm EM}+igA_{\mu}\;.
\eqa
Moreover, $m_f$ is the mass of a quark of flavor $f$
and there is a sum of flavors in Eq.~(\ref{qcdl}).
The nonabelian gauge field is $A_{\mu}=t_aA_{\mu}^a$,
$t_a={1\over2}\lambda_a$, and
$\lambda_a$ are the Gell-Mann matrices.
Finally ${\cal L}_{\rm gf}$ and ${\cal L}_{\rm ghost}$
are the gauge-fixing and ghost part of the Lagrangian, respectively.

The partition function in QCD can be written as
\bqa\nonumber
{\cal Z}&=&\int{\cal D}A_{\mu}{\cal D}\bar{\psi}_f{\cal D}\psi_f\,
e^{-S_{\rm QCD}}
\\ &=&
\int {\cal D}A_{\mu}e^{-S_g}\det(/\!\!\!\!\!\!D(B)+m_f)\;,
\eqa
where $S_{\rm QCD}$ is the Euclidean action for QCD.
In the second line, we have integrated over the fermions
which can be done exactly since ${\cal L}_{\rm QCD}$
is bilinear in the quark fields.
Moreover,
$S_g$ is the Euclidean action for the gluons
and 
\bqa
/\!\!\!\!\!\!D(B)&=&
\begin{pmatrix}
  0&iX\\
  iX^{\dagger}&0
\end{pmatrix}\;,\\ 
iX&=&D_0+i{\boldsymbol \sigma}\cdot{\bf D}\;.
\eqa
This yields
\bqa
\det\left(/\!\!\!\!\!\!D(B)+m_f\right)&=&\det\left[X^{\dagger}X+m_f^2\right]\;.
\eqa
The last equation shows that the fermion determinant is manifestly positive.
As in the case of finite isospin chemical potential, QCD in a magnetic field is also free of the sign problem, and one can therefore carry out
lattice simulations. Interestingly, the combination of finite isospin and magnetic field is free of the sign problem only if the charges of the $u$ and $d$-quark are the same. This is of course not real QCD, but it offers the possibility
to compare lattice predictions with those of low-energy effective theories and models.

In this review, we will discuss (inverse)
magnetic catalysis and the phase diagram of QCD in a strong magnetic background, 
paying attention to recent developments.
There are other reviews~\cite{review,dima,ourrev,fariasrev}
focusing on different aspects of the field.
The paper is organized as follows. In the next section, we discuss
the physics of magnetic catalysis at $T=0$. 
In Sec.~\ref{nonzero}, we introduce the Polyakov loop and discuss
magnetic catalysis in model calculations at nonzero temperature. 
In Sec.~\ref{lattice}, we review 
inverse magnetic catalysis on the lattice focusing on the competing sea and valence effects.
In Sec.~\ref{improve}, the improvement of models to 
incorporate inverse magnetic catalysis is discussed and in 
Sec.~\ref{summary}, we summarize.
In \ref{appa}, we discuss renormalization of the quark-meson model in the on-shell scheme, while in \ref{appb}, we show how the parameters of the model are fixed.

\section{Magnetic catalysis at zero temperature}
Magnetic catalysis can be defined as
\begin{enumerate}
\item
  The magnitude of a condensate is enhanced by the presence of
  an external magnetic field $B$ if the condensate already is present at vanishing field.\\
\item An external magnetic field induces symmetry breaking and a nonzero
  value of a condensate when the symmetry is intact for $B=0$.
\end{enumerate}
The condensate is the expectation value of a field, which can be
either fundamental or composite. The expectation of a scalar field
$\phi$ in low-energy models
is an example of the former, while $\bar{\psi}\psi$
is the chiral condensate in e.g. the NJL model 
or QCD is an example of the latter.
One refers to the second case as dynamical symmetry breaking by a
magnetic field. We will discuss both cases below.
The first papers on magnetic catalysis at $T=0$ appeared three decades
ago in the study of the NJL model in three dimensions~\cite{lemmer}.
Shortly thereafter in the linear sigma model~\cite{sugu}
and the NJL model in two dimensions~\cite{klim1,klim2,klim3}.
Since then it has been demonstrated in 
QED~\cite{qedgus}, 
chiral perturbation 
theory~\cite{smilga,werbos,werbos2}, in the Walecka model
in nuclear physics~\cite{andreas}, 
and also on the 
lattice, see e.g.~\cite{cherno,chern2,braguta,delia,quarkcon}.

\label{sec:2}
In this section, we will use the two-flavor
quark-meson model as an explicit
example of a low-energy effective model of QCD that displays magnetic catalysis. The Lagrangian is
\bqa\nonumber
{\cal L}&=&-{1\over2}B^2+
{1\over2}\left[(\partial_{\mu}\sigma)(\partial^{\mu}\sigma)
  +(\partial_{\mu}\pi_0)(\partial^{\mu}\pi_0)
  \right]
  \\&& \nonumber
  +D_{\mu}^*\pi^-D^{\mu}\pi^+
-{1\over2}m^2(\sigma^2+\pi_0^2+2\pi^+\pi^-)
\nonumber
\\ && \nonumber
-{\lambda\over24}(\sigma^2+{\pi}_0^2+2\pi^+\pi^-)^2+h\sigma
\\ &&
+\bar{\psi}\left[
i\gamma^{\mu}D_{\mu}
-g(\sigma+i\gamma^5{\bf\tau}\cdot{\boldsymbol\pi})\right]\psi\;,
\label{lag}
\eqa
where $D_{\mu}=\partial_{\mu}+iqA_{\mu}$ is the
covariant derivative
$\sigma$, ${\boldsymbol\pi}=(\pi_0,\pi_1,\pi_2)$ 
are the meson fields,
$\pi^{\pm}={1\over\sqrt{2}}(\pi_1\mp i\pi_2)$
, $\tau_a$ are the Pauli matrices,
$\psi$ is a color $N_c$-plet, a four-component Dirac spinor as well as a flavor doublet 
\bqa
\psi&=&
\left(
\begin{array}{c}
u\\
d
\end{array}\right)\;.
\eqa
In the absence of an abelian gauge field in 
Eq.~(\ref{lag}), the symmetry is $SU(2)_L\times SU(2)_R$
for $h=0$, otherwise it is $SU(2)_V$.
In its presence, the Lagrangian Eq.~(\ref{lag}) has a $U(1)_L\times U(1)_R$ symmetry for $h=0$, otherwise it is $U(1)_V$. The reason is that one cannot transform a
$u$-quark into a $d$-quark due to their different
electric charges. 
Defining $\Delta^{\pm}={1\over\sqrt{2}}(\sigma\pm i\gamma^5\pi_0)$, the two sets of transformations are
1) $u\rightarrow e^{-i\gamma^5\alpha}u$, $d\rightarrow e^{i\gamma^5\alpha}d$, 
$\Delta^{\pm}\rightarrow\Delta^{\pm}e^{\pm2i\gamma^5\alpha}$, and $\pi^{\pm}\rightarrow\pi^{\pm}$, and 
2) $u\rightarrow e^{i\alpha}u$, $d\rightarrow e^{-i\alpha}d$,
$\Delta^{\pm}\rightarrow\Delta^{\pm}$,
and
$\pi^{\pm}\rightarrow\pi^{\pm}e^{\pm2i\alpha}$.

After symmetry breaking,
the sigma field has a nonzero expectation value $\phi_0$. The classical potential is
\bqa
V_0&=&
{1\over2}B^2+
{1\over2}m^2\phi_0^2+{\lambda\over24}\phi_0^4-h\phi_0\;.
\eqa
The tree-level relations between the parameters of the Lagrangian
$m^2$, $\lambda$, $g$, and $h$
and the physical masses $m_{\sigma}$ and $m_{\pi}$, the pion decay constant $f_{\pi}$, and the quark mass $m_q$ are
\begin{align}
\label{tr1}
m^2&=-{1\over2}\left(m_{\sigma}^2-3m_{\pi}^2\right)\;,
&\lambda&=3{(m_{\sigma}^2-m_{\pi}^2)\over f_{\pi}^2}\;,\\
g^2&={m_q^2\over f_{\pi}^2}\;,
&h&=m_{\pi}^2f_{\pi}\;.
\label{tr4}
\end{align}
Using the relations (\ref{tr1})--(\ref{tr4}), we obtain
\bqa
V_0&=& \nonumber
{1\over2}B^2+\frac{3}{4}m_\pi^2 f_\pi^2\frac{\Delta^2}{m_q^2}
 -\frac{1}{4}m_\sigma^2 f_\pi^2\frac{\Delta^2}{m_q^2} 
+ \frac{1}{8}m_\sigma^2 f_\pi^2\frac{\Delta^4}{m_q^4}
\\ &&
- \frac{1}{8}m_\pi^2 f_\pi^2\frac{\Delta^4}{m_q^4}
-m_\pi^2f_\pi^2\frac{\Delta}{m_q}\;,
\label{classical}
\eqa
where we have introduced $\Delta=g\phi_0$.
The minimum of the classical potential is given by $\Delta=gf_{\pi}$.

The classical potential has by construction
its minimum at $\Delta=m_q$ or $\phi_0=f_{\pi}$. In the large-$N_c$ limit, 
the mesons are included at tree level, while we include the Gaussian fluctuations of the fermions. Including the one-loop corrections from the fermions  using a minimal subtraction scheme, leaves a renormalized one-loop effective
potential that depends on the renormalization scale $\Lambda$. 
The minimum of the effective potential therefore depends on $\Lambda$.
In order to ensure that the one-loop effective potential has its minimum at $\phi_0=f_{\pi}$ for zero magnetic field $B$, several methods have been used in the literature.
One method is simply to subtract the one-loop contribution to the effective potential
for $B=0$. Then the renormalization scale dependence drops out and the correction to Eq.~(\ref{classical}) is a finite $B$-dependent term that vanishes for $B=0$. However this is inconsistent since one includes fermion
fluctuations in the effective potential at finite magnetic field, but not for $B=0$. Moreover, it is also incorrect since  Eqs.~(\ref{tr1})--(\ref{tr4})
are tree-level relations that receive radiative corrections.
One can also choose a specific value for $\Lambda$ such that the one-loop correction
to the position of the minimum of the effective potential vanishes.
In this case, one has included quantum fluctuations also for $B=0$, but again, the tree-level relations between the parameters of the Lagrangian and physical
quantities receive loop corrections. 
In order to be consistent, the parameters of the Lagrangian must be determined to the same
order in the loop expansion as one calculates the effective potential. The solution to the problem is to combine
the minimal subtraction scheme with the on-shell 
scheme~\cite{sirlin,sirlin2,bohm,hollik}. 
In this way one includes loop corrections to Eqs.~(\ref{tr1})--(\ref{tr4}),
while at the same time ensures that the effective potential has its minimum at $\Delta=gf_{\pi}$. 
Details of the 
renormalization of the one-loop effective potential 
in the large-$N_c$ limit can be found in~\ref{appa}
and the parameter fixing in~\ref{appb}.
It reads
\begin{widetext}
\bqa\nonumber
V_{\rm 0+1}&=&
\frac{3}{4}m_\pi^2 f_\pi^2
\left\{1-\frac{4 m_q^2N_c}{(4\pi)^2f_\pi^2}m_\pi^2F^{\prime}(m_\pi^2)
\right\}\frac{\Delta^2}{m_q^2}
\\ \nonumber &&
 -\frac{1}{4}m_\sigma^2 f_\pi^2
\left\{
1 +\frac{4 m_q^2N_c}{(4\pi)^2f_\pi^2}
\left[ \left(1-\mbox{$4m_q^2\over m_\sigma^2$}
\right)F(m_\sigma^2)
 +\frac{4m_q^2}{m_\sigma^2}
-F(m_\pi^2)-m_\pi^2F^{\prime}(m_\pi^2)
\right]\right\}\frac{\Delta^2}{m_q^2} 
\\ \nonumber 
 & & + \frac{1}{8}m_\sigma^2 f_\pi^2
\left\{ 1 -\frac{4 m_q^2  N_c}{(4\pi)^2f_\pi^2}\left[
\frac{4m_q^2}{m_\sigma^2}
\left( 
\log\mbox{$\Delta^2\over m_q^2$}
-\mbox{$3\over2$}
\right) -\left( 1 -\mbox{$4m_q^2\over m_\sigma^2$}\right)F(m_\sigma^2)
+F(m_\pi^2)+m_\pi^2F^{\prime}(m_\pi^2)\right]
 \right\}\frac{\Delta^4}{m_q^4}
\\ && \nonumber
- \frac{1}{8}m_\pi^2 f_\pi^2
\left[1-\frac{4 m_q^2N_c}{(4\pi)^2f_\pi^2}m_\pi^2F^{\prime}(m_\pi^2)\right]
\frac{\Delta^4}{m_q^4}
-m_\pi^2f_\pi^2\left[
1-\frac{4 m_q^2  N_c}{(4\pi)^2f_\pi^2}m_\pi^2F^{\prime}(m_\pi^2)
\right]\frac{\Delta}{m_q}
\\
&&
+{1\over2}B^2
-{8N_c\over(4\pi)^2}\sum_f(q_fB)^2\left[
{\zeta^{(1,0)}(-1,x_f)}
+{1\over4}x_f^2
-{1\over2}x_f^2\log x_f+{1\over2}x_f\log x_f
-{1\over12}\log{m_q^2\over2|q_fB|}
-{1\over12}
\right]\;,
\label{fullb}
\eqa
where $x_f={\Delta^2\over2|q_fB|}$ and $\zeta(a,x)$
is the Hurwitz zeta-function. 
Here and in the remainder of the paper,
$\zeta^{(1,0)}(n,x_f)={\partial \zeta(n+\epsilon,x_f)\over\partial\epsilon}|_{\epsilon=0}$, in Eq.~(\ref{fullb}), $n=-1$.
Finally, $F(p^2)$ and $F^{\prime}(p^2)$
are defined in Eqs.~(\ref{fdef})--(\ref{fpdef}).
\end{widetext}
The first four lines of the one-loop effective potential are independent of the magnetic field and this part was first calculated in Ref.~\cite{allofus}.
The last line is the $B$-dependent correction to $V_{\rm 0+1}$.
Note also that final result is independent of the renormalization scale $\Lambda$.

\begin{figure}[htb]
\centering\includegraphics[width=0.45\textwidth]{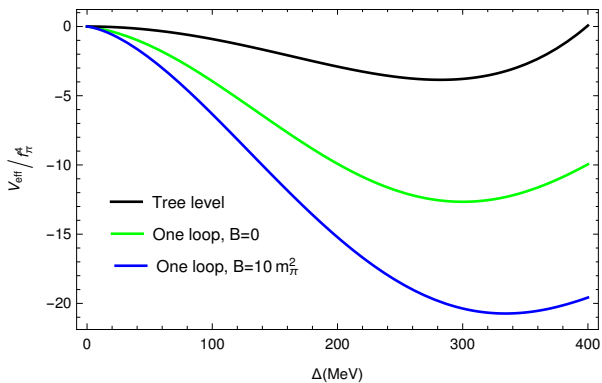}
\caption{Effective potential as a function of $\Delta$
normalized by $f_{\pi}^4$ at $T=0$. The black line is the tree-level result, the green and blue lines are the one-loop result for zero magnetic field and for
$eB=10m_{\pi}^2$. See main text for details.}
\label{cata1}
\end{figure}

In Fig.~\ref{cata1}, we show the effective potential 
divided by $f_{\pi}^4$
at $T=0$. The black
line is the tree-level potential Eq.~(\ref{classical}), while the
green and blue lines are the one-loop effective potential Eq.~(\ref{fullb}) for $|eB|=0$ and $|eB|=10 m_{\pi}^2$, respectively. 
We have used $m_{\sigma}=600$ MeV, $m_{\pi}=140$ MeV,
$f_{\pi}=93$ MeV, and $m_q=300$ MeV. 
The classical potential 
as well the one-loop
effective potential with $|eB|=0$ 
both have a minimum at $\Delta=gf_{\pi}$
by construction. Notice, however, that the latter is significantly deeper.
The blue line with $|eB|=10 m_{\pi}^2$,
shows that the minimum of the effective moves to a larger value, i. e. the system exhibits magnetic catalysis.

While the above clearly demonstrates magnetic catalysis
numerically, 
we would like to gain insight in the mechanism behind the effect.
Instead of analyzing Eq.~(\ref{fullb}), we will discuss the
gap equation in the NJL model. 

In order to simplify the discussion, we will consider the NJL model with a 
single quark flavor and color, $N_f=N_c=1$ with electric charge $q_f$.
In the chiral limit, the Lagrangian is~\cite{reg1}
\bqa
{\cal L}&=&
\bar{\psi}i\gamma^{\mu}\partial_{\mu}
\psi
+{G\over2}\left[(\bar{\psi}\psi)^2+(\bar{\psi}i\gamma^5\psi)^2\right]
\;.
\label{njllag}
\eqa
This Lagrangian has a $U(1)_V\times U(1)_A$ symmetry.
We introduce the gap $M=-G\langle\bar{\psi}\psi\rangle$
and linearize the interaction terms, writing 
$(\bar{\psi}\psi)^2\approx\langle\bar{\psi}\psi\rangle^2+2\langle\bar{\psi}\psi\rangle\bar{\psi}\psi$ and $(\bar{\psi}i\gamma^5\psi)^2\approx0$.
$M$ is now an effective quark mass arising after breaking the axial symmetry
spontaneously, i.e. when $\langle\bar{\psi}\psi\rangle\neq0$.
In the mean-field approximation, we perform the Gaussian integral over
the fermion field giving rise to the following one-loop effective
potential,
\bqa
V_{0+1}&=&{M^2\over2G}-2\int{d^4p\over(2\pi)^4}\log\left[p^2+M^2\right]\;.
\eqa
The gap equation for $M$ is found by extremizing $V_{0+1}$ which yields
\bqa
{M\over4G}&=&M\int{d^4p\over(2\pi)^4}
{1\over p^2+M^2}\;.
\label{gapleik}
\eqa
Conventionally, since the NJL model is non-renormalizable,
one has used a three-dimensional or a four-dimensional momentum cutoff $\Lambda$ to regulate divergences. 
If $\Lambda$ is a four-dimensional cutoff, the gap equation (\ref{gapleik})
reads for $M\ll\Lambda$
\bqa
M\left[{4\pi^2\over G}-\Lambda^2+M^2\log{\Lambda^2\over M^2}\right]&=&0\;.
\eqa
$M=0$ is always a solution, however for $G>G_c={4\pi^2\over\Lambda^2}$ there 
is also a nontrivial solution. Thus for $G$ larger than the critical
value ${4\pi^2\over\Lambda^2}$, 
quantum fluctuations induce symmetry breaking in the model.

At finite magnetic field, the partial derivative in Eq.~(\ref{njllag})
is replaced by the covariant derivative and we add a term ${1\over2}B^2$
to the effective potential. The gap equation  becomes
\bqa
{M\over2G}&=&M{|q_fB|\over2\pi}
\sum_{s=\pm1}\sum_{k=0}^{\infty}\int{d^2p_{\parallel}\over(2\pi)^2}
{1\over p_0^2+p_z^2+M_B^2}\;,
\label{gap2}
\eqa
where $M_B^2=M^2+|q_fB|(2k+1-s)$,  $p_{\parallel}^2=p_0^2+p_z^2$
and $q_f$ is the charge.
The divergences in Eq.~(\ref{gap2}) can be isolated by
adding and subtracting the right-hand side of Eq.~(\ref{gapleik}).
The right-hand side of Eq.~(\ref{gap2}) minus the subtracted term is finite and is conveniently evaluated using dimensional regularization
in the same way as done appendix A~\cite{menes,farias}.
Finally, we impose a four-dimensional cutoff on the added term
as in Eq.~(\ref{gapleik}). 
Factoring out the trivial solution $M=0$,
this yields the regularized gap equation
\bqa\nonumber
{4\pi^2\over G}-\Lambda^2+M^2\log{\Lambda^2\over M^2}
-{2|q_fB|}\left[\zeta^{(1,0)}(0,x_f)+x_f
\right. \\ \left.
-{1\over2}(2x_f-1)\log{x_f}\right]&=&0\;,
\eqa
where $x_f={M^2\over2|q_fB|}$.
This equation has only a nonzero $M$ as solution. For $G<G_c$, the
solution is~\cite{reg1,gus2}
\bqa
M^2&=&{|q_fB|\over\pi}\exp\left[{-{1\over|q_fB|}}\left({4\pi^2\over G}-\Lambda^2\right)\right]\;.
\label{sol}
\eqa
In the limit $|q_fB|\rightarrow0$, this solution connects to the trivial solution $M=0$. In the lowest Landau level approximation, the
gap equation has solution $M^2=\Lambda^2e^{-4\pi^2/G|q_fB|}$, which is reminiscent of Eq.~(\ref{sol}) if we identify the cutoff
$\Lambda$ with $\sqrt{|q_fB|}$. We can then think of magnetic catalysis
as a 1+1 dimensional phenomenon, i.e. a dimensional reduction from 3+1 dimensions has taken place. The functional form of the gap equation
is the same as for the gap equation in BCS theory of superconductivity
as well as the gap equation found in the large-$N$ limit of
$O(N)$-symmetric nonlinear sigma model in 1+1 dimensions.
The 1+1 dimensional nature of magnetic catalysis raises
the question of whether this phenomenon is in conflict
with the Coleman theorem, which forbids spontaneous symmetry
breaking in less than two spatial dimensions at zero 
temperature~\cite{coleman}. As pointed out in Ref.~\cite{gus2},
the field $\bar{\psi}\psi$ is neutral with respect to the
magnetic field. The neutral pion is the associated
Goldstone boson that appears after breaking the $U(1)$
symmetry. The charged pions are now massive even in the
chiral limit.

There are other ways of regularizing the gap equation~(\ref{gap2})
or the fermion contribution to the one-loop effective potential
(\ref{deter}), for example Schwinger's proper time method~\cite{swing}. 
Let us illustrate this by computing the corresponding
bosonic functional determinant, which shows up in chiral perturbation
theory.
It is based on the representation in Euclidean space
\bqa\nonumber
V_1&=&\log\det(-D_{\mu}D^{\mu}+m^2)
\\ \nonumber
&=&-\int_0^{\infty}{ds\over s}{\rm Tr}\,e^{-s(-D_{\mu}D^{\mu}+m^2)}
\\
&=&
-{|qB|\over2\pi}\sum_{k=0}^{\infty}\int_{p_{\parallel}}
\int_0^{\infty}{ds\over s}
e^{-s(p_{\parallel}^2+|qB|(2k+1)+m^2)}\;,
\eqa
where the sum over Landau levels $k $ as well the momentum integral
over $p_{\parallel}$ is convergent. The result 
is
\bqa
V_1&=&-{(e^{\gamma_E}\Lambda^2)^{\epsilon}\over(4\pi)^2}
\int_0^{\infty}{ds\over s^{3-\epsilon}}e^{-m^2s}
{|qB|s\over\sinh(|qB|s)}\;.
\label{div22}
\eqa
For $\epsilon=0$,
the integral is divergent for small $s$, i.e. for large momentum. 
By adding and subtracting the divergent terms, we can isolate the
divergences. One finds
\bqa \nonumber
V_1&=&-{(e^{\gamma_E}\Lambda^2)^{\epsilon}\over(4\pi)^2}
\int_0^{\infty}{ds\over s^{3-\epsilon}}e^{-m^2s}
+{(qB)^2\over6(4\pi)^2}\int_0^{\infty}{ds\over s^{1-\epsilon}}e^{-m^2s}
\\ &&\nonumber
-{(e^{\gamma_E}\Lambda^2)^{\epsilon}\over(4\pi)^2}\int_0^{\infty}{ds\over s^{3-\epsilon}}e^{-m^2s}
\left[{|qB|s\over\sinh(|qB|s)}-1+{(qBs)^2\over6}\right]\;.
\\ &&
\label{swing}
\eqa
The integrals in the first line are divergent for $\epsilon=0$, while the last integral is convergent. The divergences show up as poles in $\epsilon$. 
The first term in Eq.~(\ref{swing}) is a vacuum energy counterterm
while the second term corresponds to 
charge and wavefunction renormalization~\cite{swing}.
The last integral in Eq.~(\ref{swing}) can be calculating
exactly and involves the Hurwitz zeta function.
Using the proper time method with the momentum integrals 
evaluated in $d=2-2\epsilon$ dimensions yields the same results
as those obtained by combining dimensional regularization 
and zeta-function regularization, as done in \ref{appa}.
Alternatively, one can evaluate Eq.~(\ref{swing})
with $\epsilon=0$ using a cutoff $1/\Lambda^2$
as the lower limit of the $s$-integration in the divergent integrals.

The regularization methods discussed so far separates in clean
way the $B$-independent divergences from the $B$-dependent
terms whether they are finite or 
divergent.~\footnote{They are finite in the gap equation
Eq.~(\ref{gap2}), but divergent in the effective potential, cf.
Eq.~(\ref{swing}).} 
There are other regularization methods that do not separate
this contributions, for example a sharp cutoff imposed
directly on the integral in Eq.~(\ref{div22})
or a form factor that is a function of e.g. $p_z^2+2k|qB|(2k+1-s)$.
One has to be careful choosing such
regulators since nonphysical oscillations may 
result~\cite{farreg0,farreg}.

The on-shell scheme used to obtain the final result Eq.~(\ref{fullb})
has two important virtues as first pointed out in Ref.~\cite{hrg}.
By considering the small-$B$ (large-$x_f$) behavior of the Hurwitz zeta-function, one finds that the only contributions 
at order $B^2$ comes from
the renormalized magnetic field term ${1\over2}B^2$,
\bqa\nonumber
{\zeta^{(1,0)}(-1,x_f)}
&=&-{1\over4}x_f^2
+{1\over2}x_f^2\log x_f-{1\over2}x_f\log x_f
\\ &&
+{1\over12}\log x_f
+{1\over12}+{1\over720}{1\over x_f^2}+...\;.
\eqa
It also ensures that the magnetic-field contribution 
to the effective potential and the
magnetization vanishes in the limit $m_q\rightarrow\infty$.
Both properties are expected from a physical point of view.
It leads to a paramagnetic vacuum, in agreement with 
the hadron resonance gas model calculations~\cite{hrg} and
lattice QCD lattice simulations~\cite{para}. 
Other renormalization schemes, such as the (modified) minimal
subtraction scheme
are connected to the above by a finite renormalization. However,
the effective potential and the magnetization then grow logarithmically in the 
limit  $m_q\rightarrow\infty$.
Similar remarks apply to the (P)NJL model. 
Regularizing the model by imposing a UV cutoff $\Lambda$
on the divergent integrals in the fermionic version of Eq.~(\ref{swing}),  the authors of~\cite{sidney} show that it predicts
diamagnetic behavior for low values of $B$
and paramagnetic behavior for large magnetic fields.
By defining a subtraction procedure that resembles the 
renormalization of the magnetic field in the on-shell scheme, 
their effective potential leads to paramagnetic behavior
as seen on the lattice.

After having discussed magnetic catalysis in models, we now turn
to lattice gauge theory.
The first lattice simulations were carried out for an $SU(2)$ gauge group
for magnetic field strengths up to $\sqrt{eB}\sim 3$ GeV
in the quenched approximation, i.e. setting
the quark determinant to unity~\cite{cherno}.
The simulations confirmed that the quark condensates is enhanced by the magnetic field and that the enhancement is qualitative linear with $eB$.
The quark condensate itself was calculated using the Banks-Casher relation~\cite{banks}, which relates the density of eigenvalues close to 
zero of the Dirac operator and the condensate. Their calculations showed
a monotonic increase of the spectral density for typical gauge field configurations. This enhancement induced by the magnetic field can be considered
the basic mechanism behind magnetic catalysis. 
Below, we will discuss this mechanism further, here it suffices to add that the
enhancement of the spectral density as a function of $B$ for typical gauge configuration is also seen in full QCD~\cite{bruck}.
Even in the free case, there is a proliferation of small eigenvalues due to the degeneracy of states, which in a constant magnetic field
is proportional to $|qB|$~\cite{smilga}.

\section{Magnetic catalysis at nonzero temperature}
\label{nonzero}
In the previous section, we reviewed magnetic catalysis at $T=0$ in some detail.
A survey of the literature shows that it is a robust feature of lattice simulations
as well model calculations: Magnetic catalysis does not depend on particular values of the masses or couplings. Since the condensate increases as a function of the
magnetic field, it should raise the transition temperature for the chiral transition.
This expectation was made explicit a long time ago in Ref.~\cite{smilga}.
A number of model calculations have confirmed this expectation, e.g.~\cite{lemmer,fraga0,fraga,gatto,gatto2,skokov,marco1,william}, although
some of them suggest that the chiral and deconfinement
transition split for larger values of $eB$.

\begin{figure}[htb]
\centering
\includegraphics[width=0.45\textwidth]{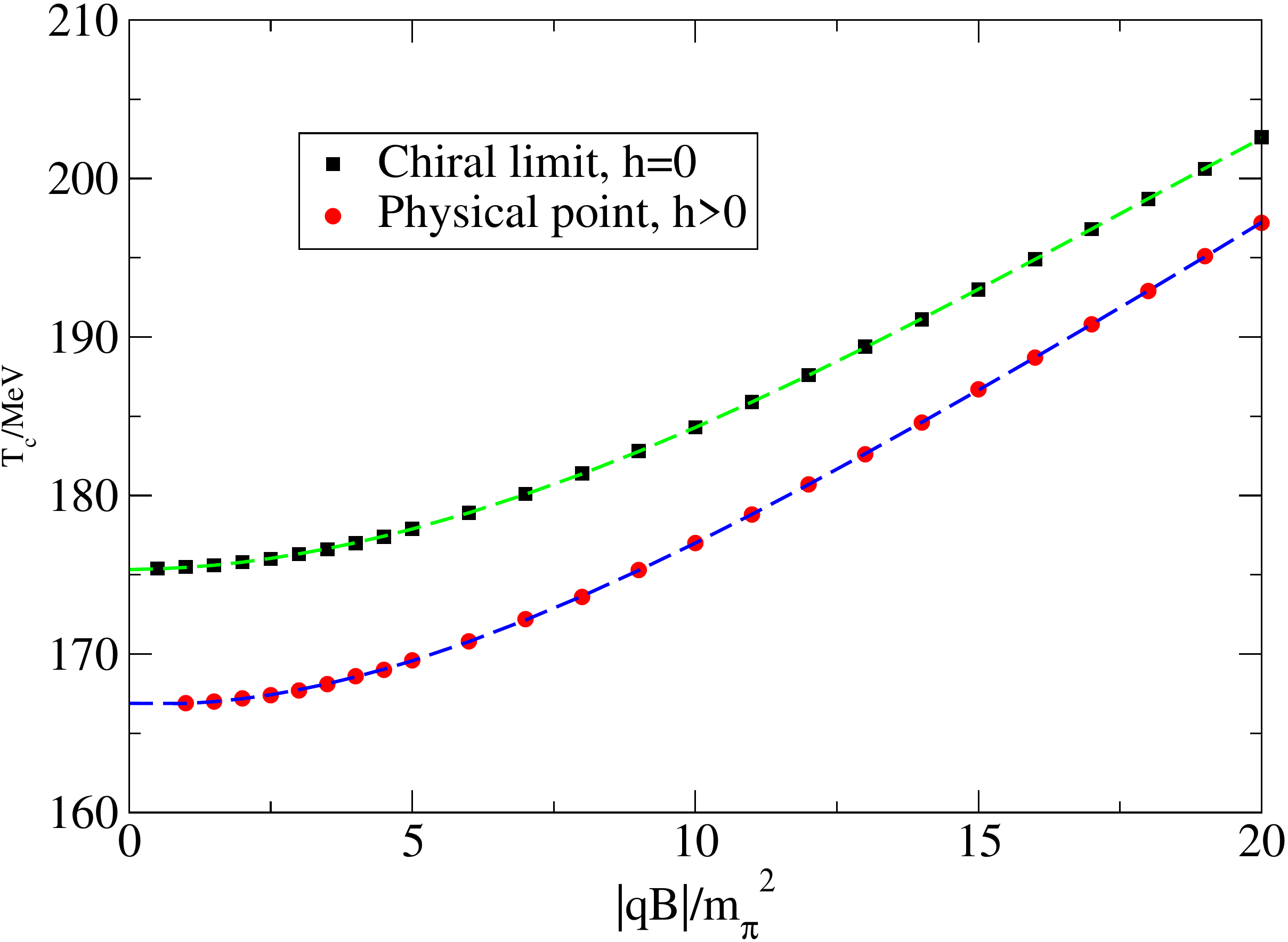}
\caption{$T_{pc}(B)$ as a function of $|qB|$ in units of $m_{\pi}^2$ in the quark-meson model. The green points are in the chiral limit and the red points are at the
physical point. See main text for details.
Figure taken from Ref.~\cite{ourrev}.}
\label{cata00}
\end{figure}

The effective potential of the quark-meson
model in the large-$N_c$ approximation at finite temperature is
\bqa\nonumber
V_{0+1}^T&=&V_{0+1}-
\sum_f{2N_c|q_fB|\over2\pi}T\sum_{s=\pm1}
\sum_{k=0}^{\infty}
\int_{p_{z}}
\log\left[1+e^{-\beta E_f}\right]\;,
\\ &&
\label{tempcont}
\eqa
where $E_f=\sqrt{p_z^2+\Delta^2+|q_fB|(2k+1-s)}$ and $V_{0+1}$ is given by Eq.~(\ref{fullb}).
In Fig.~\ref{cata00}, we show the results of a typical calculation where the
quark-meson model was used. 
The curves show the transition temperature for the chiral transition as a function of
$|qB|/m_{\pi}^2$ in the chiral limit (green points) and at the physical point (red points). At $B=0$, the gap between the two critical temperatures is approximately
10 MeV, which decreases as $|qB|$ grows. In both cases, it is clear that
the transition temperature increases with the magnetic field. 
Here the transition temperature was defined as the inflection point of the
curve $\phi_0(T)$ at the physical point and $\phi_0(T)=0$ in the chiral limit.
An alternative definition of the critical temperature is the peak of the chiral susceptibility
~\footnote{Using the peak of ${\partial \Phi\over\partial T}$, where $\Phi$ is the Polyakov loop, yields a transition temperature for deconfinement, which is very close to the chiral transition temperature.} 
\bqa
\chi&=&{\partial\langle\bar{\psi}\bar{\psi}\rangle\over\partial T}\;.
\eqa


In QCD, two transitions take place as one increases the temperature, namely the chiral transition and the deconfinement transition.
Lattice calculations suggest that chiral symmetry is “restored” at a temperature of approximately $T^{\chi}_c$= 155 MeV~\cite{aoki,aoki2,borsa,baza0,baza}
though strictly speaking the transition is only a crossover. The crossover temperature is defined by the peak of the chiral susceptibility. This temperature is slightly less than the crossover temperature for the deconfinement transition, $T_c^{\rm dec}=$170 MeV. However this temperature difference is observable dependent. In most cases,  $T_c^{\rm dec}$ has been determined by the behavior of the Polyakov loop. Recently, it has been defined by the behavior of the  quark  entropy  and  in  this  case  the two  crossover temperatures agree within errors~\cite{baza}.

We will next discuss the Polyakov loop and how it can be
incorporated in model calculations.
The Wilson line is defined as
\bqa
L({\bf x})&=&{\cal P}\exp\left[
i\int_0^{\beta}d\tau A_4({\bf x},\tau)
\right]\;,
\label{wilson}
\eqa
where ${\cal P}$ denotes path ordering, $A_4=iA_0$ and $A_0=t_aA_0^a$. 
The Polyakov loop operator $l$ is the trace of the Wilson line~(\ref{wilson}).
Together with its Hermitian conjugate, it is defined as
\begin{align}
l&={1\over N_c}{\rm Tr}L\;,&
l^{\dagger}={1\over N_c}{\rm Tr}L^{\dagger}\;,
\end{align}
where $N_c$ is the number of colors. 
The expectation values of $l$ and $l^{\dagger}$
are denoted by $\Phi$ and $\bar{\Phi}$.
Under the center group
$Z_{N_c}$ of the gauge group $SU(N_c)$,
the Polyakov loop transforms as 
$\Phi\rightarrow e^{2\pi in\over N_c}\Phi$ with $n=0,1,2...N_c-1$.
In pure-glue QCD it is an order parameter for confinement, while for QCD with dynamical fermions it is only an approximate order parameter~\cite{yaffe}. Note also that $\Phi=\bar{\Phi}$
at zero density, i.e. for $\mu_f=0$.

For $N_c=3$ and in the Polyakov gauge, one can write a nonabelian background gauge field as
\bqa
A_4&=&t_3A_4^{3}+t_8A_4^{8}\;.
\eqa
Introducing the fields $\phi_1={1\over2}\beta A_4^{3}$
and $\phi_2={1\over2\sqrt{3}}\beta A_4^{8}$, the thermal Wilson
line reads for constant gauge fields
\bqa
L&=&
\left(
\begin{array}{ccc}
e^{i(\phi_1+\phi_2)}&0&0\\
0&e^{i(-\phi_1+\phi_2)}&\\
0&0&e^{-2i\phi_2}\\
\end{array}\right)\;.
\eqa
Since the Polyakov loop is an approximate order parameter for deconfinement, 
the strategy put forward in Ref.~\cite{polyakov1} is to write down
a phenomenological effective potential for $\Phi$, $\bar{\Phi}$
and the chiral condensate
that describes the thermodynamics of the system.
This potential consists of a gluonic part $U(\Phi,\bar{\Phi})$ as well as a matter part. 
The term $U(\Phi,\bar{\Phi})$ is constructed
such that it reproduces the pure-glue pressure calculated on the 
lattice~\cite{glue}.
A number of different forms of $U(\Phi,\bar{\Phi})$ have been proposed~\cite{ratti,sjafer,ratti2,ratti3}.
In Ref.~\cite{ratti}, they used a Polynomial expansion incorporating
the $Z_3$ center symmetry,
\bqa\nonumber
{{ U}\over T^4}&=&-{1\over2}b_2(T)\Phi\bar{\Phi}
-{1\over6}b_3(T)\left[
\Phi^3+\bar{\Phi}^3\right]+{1\over4}b_4(\Phi\bar{\Phi})^2\;.
\\ &&
\eqa
Here the coefficients are 
\bqa\nonumber
b_2(T)&=&6.75 -1.95\left({T_0\over T}\right)
+2.625\left({T_0\over T}\right)^2
-7.44\left({T_0\over T}\right)^3\;,
\\ &&
\\
b_3&=&0.75\;,\\
b_4&=&7.5\;,
\eqa
and $T_0=270$ MeV, the transition temperature for pure-glue QCD~\cite{glue}.
A drawback of the proposed pure-glue potentials is that they are independent of the number of flavors $n_f$. The transition temperature for $B=0$ depends on the number of
flavors
and one should incorporate the back-reaction from the
fermions to the gluonic sector~\cite{sjafer}. This is done by using an $n_f$-dependent $T_0$. Once the coupling between the  gluonic sector and the matter sector has been implemented, the
two transitions take place at approximately the same temperature: The chiral
transition moves to larger temperatures, while the deconfinement transition moves to lower temperatures.
Finally, the Polyakov-loop potential is coupled to the matter
sector via replacing the partial derivatives in the 
fermionic part of the Lagrangian by covariant ones including the
constant background gauge field.
This is implemented by making the substitution
\bqa\nonumber
\log\left[1+e^{-\beta E_f}\right]
&\rightarrow&
{1\over6}\log\left[
1+3\Phi e^{-\beta E_f}+3\bar{\Phi}e^{-2\beta E_f}
\right.\\ &&\left. \nonumber
+e^{-3\beta E_f}\right]
\\ &&
\hspace{-2cm}
+
{1\over6}\log\left[
1+3\bar{\Phi} e^{-\beta E_f}+3{\Phi}e^{-2\beta E_f}+e^{-3\beta E_f}\right]
\eqa
in Eq.~(\ref{tempcont}).
In the same way, the Fermi-Dirac distribution function is
generalized, 
\bqa
n_F(\beta E_f)&=&{1+2\bar{\Phi}e^{\beta E_f}+\Phi e^{2\beta E_f}
\over1+3\bar{\Phi}e^{\beta E_f}+3\Phi e^{2\beta E_f}+ e^{3\beta E_f}}\;.
\label{genfer}
\eqa
For small values of the Polyakov loop, $\Phi\approx0$,
$\bar{\Phi}\approx0$
Eq.~(\ref{genfer}) reduces to a Fermi-Dirac distribution with
excitation energy $3E_f$, i.e. that of three quarks.
For large temperatures, when $\Phi\approx1$, $\bar{\Phi}\approx1$,
the excitation energy is $E_f$, which is the distribution function of deconfined quarks.

To the best of our knowledge, there are no systematic studies of the transition temperature as a function of the magnetic field $B$ in various
approximations. However, some interesting results 
using the quark-meson model exist. 

\begin{figure}[htb]
\centering\includegraphics[width=0.45\textwidth]{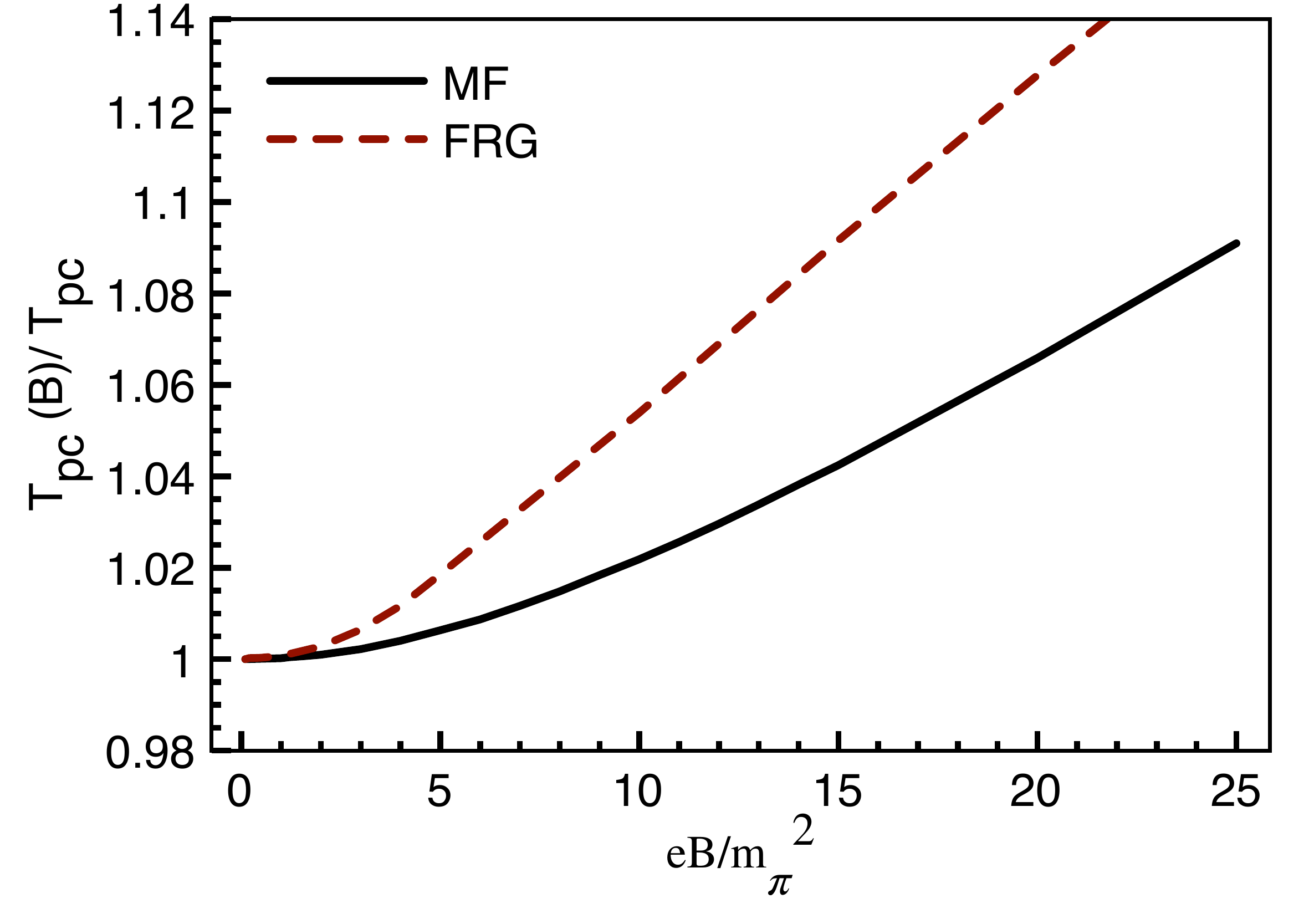}
\caption{$T_{\rm pc}(B)/T_{\rm pc}$ as a function of $eB/m_{\pi}^2$
in the quark-meson model.
Solid line is the mean-field result and the dashed line is the
result from the functional renormalization group. 
See main text for details. Figure taken from Ref.~\cite{skokov}.}
\label{cata2}
\end{figure}

Fig.~\ref{cata2}
shows the normalized transition temperature 
from Ref.~\cite{skokov} in two approximations
using the functional renormalization 
group (FRG)~\cite{wetterich}.
In this approach, one solves a flow equation for the 
effective potential numerically by lowering a sliding scale $k$
from an initial UV cutoff $k=\Lambda$ (where the effective potential is equal to the classical potential) down to $k=0$. The bare parameters at $k=\Lambda$ are tuned such that  one obtains the physical values of the masses and the pion decay constant in the vacuum (i. e. for $k=0$).
In this way, all quantum and thermal fluctuations are included. The black solid line is the mean-field result, i.e. the bosons
are excluded from the flow equation, whereas the brown
line is the result using the functional renormalization group.
Clearly, the addition of bosonic fluctuations increases the 
transition temperature significantly. 

\begin{figure}[htb]
\centering\includegraphics[width=0.50\textwidth]{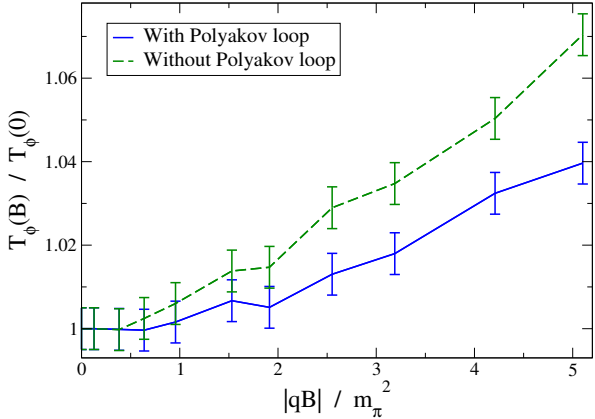}
\caption{Normalized chiral transition temperature $T_{\phi}(B)$ as a function of $|qB|$ in units of $m_{\pi}^2$ in the quark-meson model. The green points are without the Polyakov loop and the blue  points are with the
Polyakov loop. See main text for details.
Figure taken from Ref.~\cite{william}.}
\label{cata3}
\end{figure}

In Fig.~\ref{cata3}, we show the transition temperature at the physical point using the functional renormalization group~\cite{william}. 
The green points are the results without the Polyakov loop, whereas the blue points are the results including it. Clearly, the Polyakov loop lowers
the transition temperature for fixed $B$, but it is still increasing as we increase the magnetic field.
The above FRG results are obtained in the so-called local-potential approximation.
In Ref.~\cite{kamikado}, the authors added the effects of wavefunction renormalization and the curve for the critical temperature lies between
mean-field  and the local-potential approximation.
Thus the coupling of the Polyakov loop to the chiral sector is not sufficient to reproduce (qualitatively) the results seen on the lattice.

\section{(Inverse) Magnetic catalysis on the lattice}
\label{lattice}
After having discussed magnetic catalysis in low-energy models and 
theories of QCD, we next consider QCD lattice simulations.
In the past decade, there have been a number of lattice calculations of QCD in a magnetic field
~\cite{cherno,chern2,braguta,bali0,quarkcon,bruck,delia,delia1,ani,delia2,heavypi,ding0,ding1}, which have improved our understanding of QCD in a magnetic background.

\begin{figure}[htb]
\centering\includegraphics[width=0.45\textwidth]{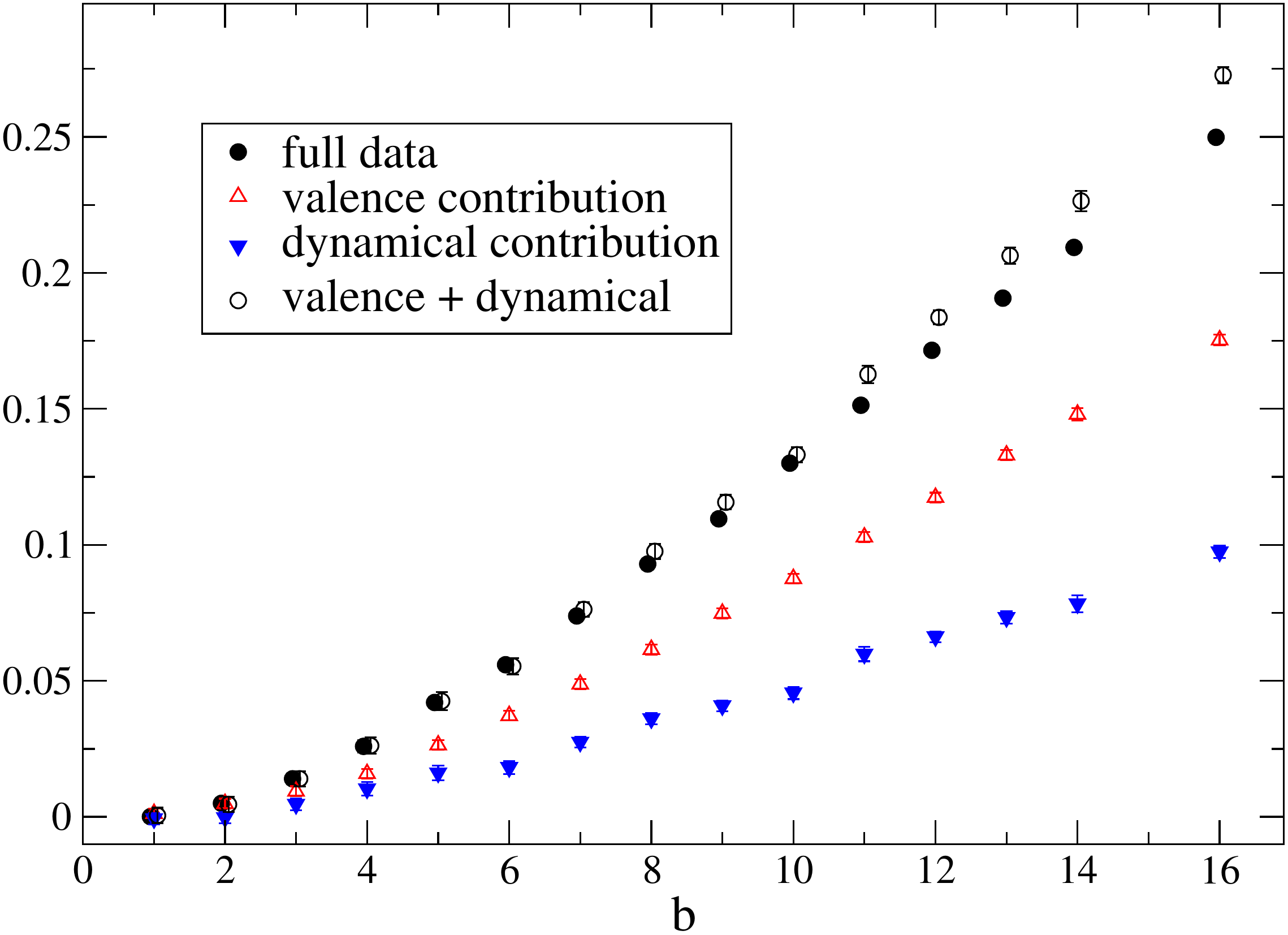}
\caption{Relative increment of the average of the $u$ and $d$
quark condensates as a function of  $b$.
Valence (red points) and dynamical (sea) (blue points) contributions, the sum of them (open circles), and the full quark condensate as a function of the dimensionless quantity $b$.
See main text for details.
Figure taken from Ref.~\cite{delia}.}
\label{cata0}
\end{figure}

In order to discuss (inverse) magnetic catalysis as seen on the lattice, it is advantageous to take a look at the path-integral representation of a number of expectation values. The QCD Lagrangian is bilinear in the quark fields 
$\psi_f$ and so one can integrate over them, giving for the partition function as a path integral over gauge configurations
$A_{\mu}$
\bqa
{\cal Z}(B)&=&\int d{A}_{\mu}
e^{-S_g}\det(D\!\!\!\!\!/(B)+m)\;,
\eqa
where $S_g$ is the Euclidean gluon action and 
$\det(D\!\!\!\!\!/(B)+m)$ is the fermion functional determinant
(suppressing flavors). 
The operator $D\!\!\!\!\!/(B)$ contains the nonabelian gauge field, which we have suppressed, as well as the abelian background $B$ that we have indicated. The quark condensate is given by
\bqa
\nonumber
\langle\bar{\psi}\psi\rangle
&=&{\partial \over\partial m}\log{\cal Z}(B)
\\ \nonumber
&=&{1\over{\cal Z}(B)}\int d{A}_{\mu}
e^{-S_g}\det(D\!\!\!\!\!/(B)+m)
{\rm Tr}(D\!\!\!\!\!/(B)+m)^{-1}\;.
\\ &&
\eqa
We can think of $P={1\over{\cal Z}(B)}e^{-S_g}\det(D\!\!\!\!\!/(B)+m)$ as a measure that depends on the gauge-field
configuration $A_{\mu}$, the magnetic field, and the quark masses. Note that the $B$-dependence is in the functional determinant as well as the the trace of the propagator. In order to study the contributions to the quark condensate coming separately from the change of the operator and the change
of the measure, it is convenient to introduce the valence and sea contributions defined as
\bqa\nonumber
\langle\bar{\psi}\psi\rangle^{\rm val}
&=&{1\over{\cal Z}(0)}\int d{A}_{\mu}
e^{-S_g}\det(D\!\!\!\!\!/(0)+m)
{\rm Tr}(D\!\!\!\!\!/(B)+m)^{-1}\;,
\\ &&
\label{valence}
\\
\nonumber
\langle\bar{\psi}\psi\rangle^{\rm sea}
&=&{1\over{\cal Z}(B)}\int d{A}_{\mu}
e^{-S_g}\det(D\!\!\!\!\!/(B)+m){\rm Tr}
(D\!\!\!\!\!/(0)+m)^{-1}\;.
\\ &&
\label{sea}
\eqa
This can be thought of as an expansion of the quark condensate around $B=0$. A priori, the sum of the two contributions needs not add up to the total quark condensate unless we are at small fields. However, it turns out that  
writing the condensate as a sum of the valence and sea contribution is remarkably good. 
This is clearly demonstrated in Fig.~\ref{cata0} from Ref.~\cite{delia}, which shows the relative increment $r$ of the 
valence and sea contributions,
their sum as well as the complete results for the 
quark condensate as a function of a dimensionless quantity $b$.
The relative increment is defined as
\bqa
r&=&{\langle\bar{\psi}\psi\rangle_B\over\langle\bar{\psi}\psi\rangle}-1\;,
\eqa
where $\langle\bar{\psi}\psi\rangle$ is the
average of the $u$ and $d$ quark condensates.
Within error, the additivity is confirmed for values of 
$b$ up to 8, which corresponds to magnetic fields up to
$eB=(500{\rm MeV})^2$~\cite{delia}. It is also of interest
to notice that both contributions work in the same direction, namely to increase the quark condensate as $B$ grows. This is unlike what happens at temperatures around the critical temperature $T_c$, as we shall see below.
As pointed out in Ref.~\cite{bruck}, $\langle\bar{\psi}\psi\rangle^{\rm sea}$
can be thought of as the quark condensate of an electrically neutral fermion flavor coupled to an electrically charged fermion flavor, since the magnetic field only
appears in the functional determinant and not in the propagator.
On the other hand, $\langle\bar{\psi}\psi\rangle^{\rm val}$ is reminiscent of the expression of the quark condensate in model calculations, except in models one does not integrate over gauge-field configurations.

Let us now turn to finite temperature. {\it Inverse magnetic catalysis} seems to have two somewhat different meanings in the
literature. The first meaning corresponds directly to the concept
magnetic catalysis discussed above: it simply means that a
condensate, for example $\langle\bar{\psi}\psi\rangle$,
decreases with the magnetic field at a fixed temperature.
The second meaning is that the transition temperature 
itself is a decreasing function of the magnetic field.

\begin{figure}[htb]
\centering\includegraphics[width=0.45\textwidth]{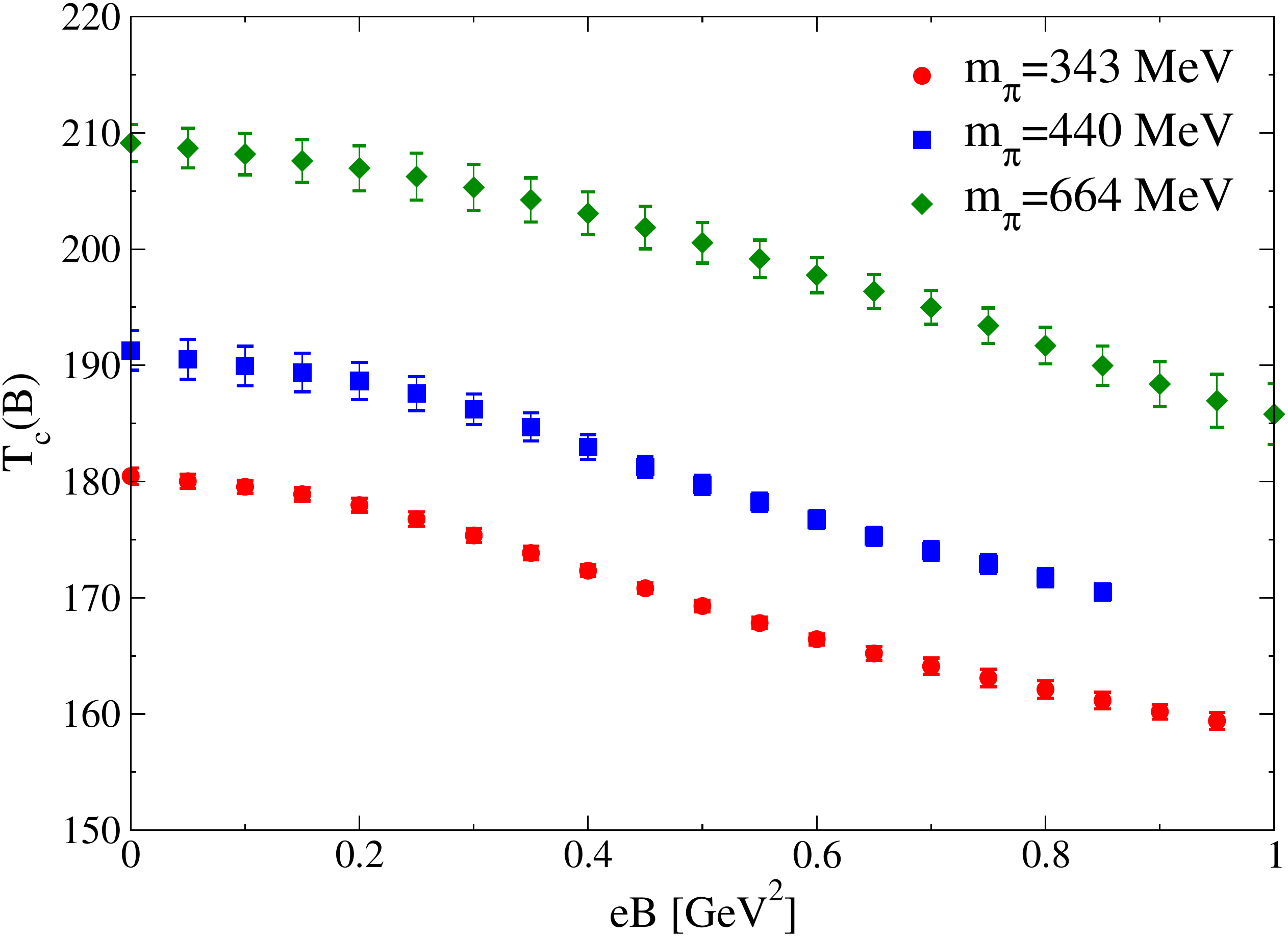}
\caption{Transition temperature in GeV for the chiral transition 
as a function of $eB$ for different values of the pion mass. See main text for details.
Figure taken from Ref.~\cite{delia2}.}
\label{invcata0}
\end{figure}

The first finite-temperature lattice simulations 
were carried out in~\cite{cherno,chern2} for $SU(2)$ gauge theory
in the quenched approximation, focusing on the $B$-dependence
of the chiral condensate for temperatures below the transition.
In two-flavor QCD, simulations at finite temperature were
carried out for pion masses in the range $200-480$ MeV
in Ref.~\cite{delia} and it
was concluded that the chiral and deconfinement transitions
take place at the same temperature and that they increase 
slightly with the external magnetic field. 
The increase of the transition temperature with $B$
is, at least qualitatively, in agreement with model calculations.
Bali et al~\cite{quarkcon,bali0} carried out 
lattice simulations at the physical point with 2+1 flavors, i.e. for quark
masses that correspond to $m_{\pi}=140$ MeV, and the result
was somewhat surprising: The transition temperature
turned out to be decreasing as $B$ increases. 
The different behavior of $T_c$ is not a consequence of the
different pion masses, rather it results from lattice artefacts
and that the results of ~\cite{delia} were not continuum 
extrapolated. Today there is consensus that the chiral
transition temperature is a decreasing function of the magnetic
field. This behavior is illustrated in 
Fig.~\ref{invcata0}, which shows the results of a recent lattice simulation~\cite{delia2},
namely the transition temperature in MeV as a function of the magnetic field $eB$
in GeV for three different pion masses. The pion mass is 343 MeV (red points), 440 MeV (blue points), and 
664 MeV (green points), which is much larger than the physical pion mass of 140 MeV. The transition temperature increases as a function of the pion mass for fixed value of
$B$, which is also known from $B=0$ calculations.

\begin{figure}[htb]
\centering\includegraphics[width=0.45\textwidth]{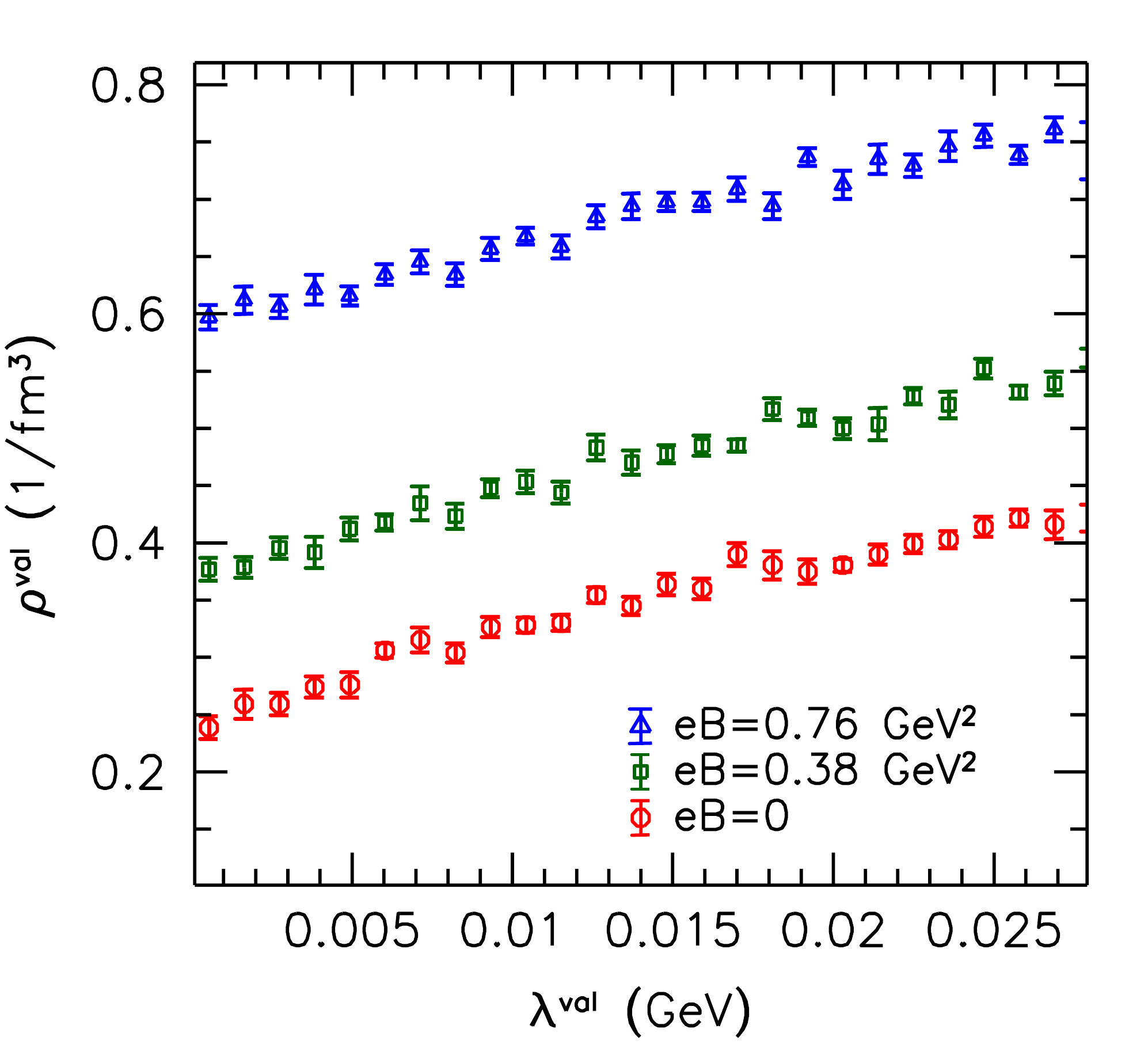}
\caption{Spectral density of the Dirac operator for three different
values of the magnetic field.
See main text for details.
Figure taken from Ref.~\cite{bruck}.}
\label{tettleik}
\end{figure}

In Ref.~\cite{bruck}, the authors carried out a thorough analysis
of the quark condensate around the critical temperature to understand the behavior of the transition temperature, focusing
on disentangling the valence and sea effects.
The valence contribution Eq.~(\ref{valence}) can also 
be written as
\bqa
\langle\bar{\psi}\psi\rangle^{\rm val}&=&\langle
{\rm Tr}(D\!\!\!\!\!/(B)+m)^{-1}\rangle_0\;,
\label{valcon}
\eqa
where the subscript indicates that the quark determinant
is without a magnetic field. The spectral density of the
quark operator for different values of the magnetic field
is shown in Fig.~\ref{tettleik}. From the figure, it is 
evident that there is an increase in the spectral density
around zero
with increasing magnetic field. The corresponding ensemble 
was generated at finite temperature, $T=142$ MeV and for vanishing magnetic background~\cite{bruck}. 
The Banks-Casher relation~\cite{banks} then implies
an increase of the valence contribution.

\begin{figure}[htb]
\centering\includegraphics[width=0.45\textwidth]{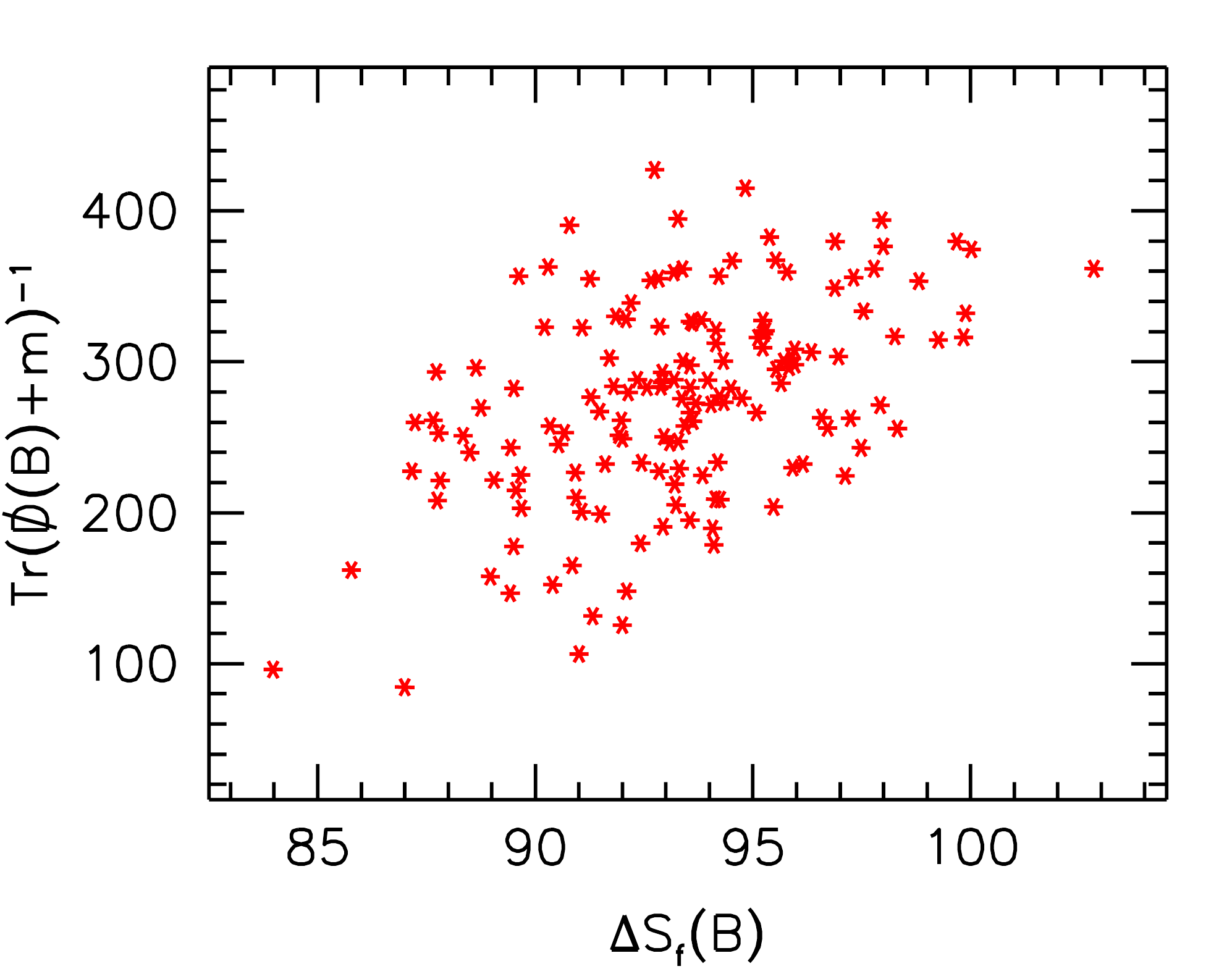}
\caption{
Scatter plot of the down-quark condensate as a function of
$\Delta S_f(B)$.
See main text for details.
Figure taken from Ref.~\cite{bruck}.}
\label{scatter}
\end{figure}

\noindent
Defining the quantity 
\bqa
-\Delta S_f(B)=\log\det(D\!\!\!\!\!/(B)+m)-\log\det(D\!\!\!\!\!/(0)+m)\;,
\eqa
the full condensate can be written as
\bqa
\langle\bar{\psi}\psi\rangle&=&
{\langle e^{-\Delta S_f(B)}{\rm Tr}(D\!\!\!\!\!/(B)+m)^{-1}\rangle_0\over\langle e^{-\Delta S_f(B)}\rangle_0}\;.
\label{fullcon}
\eqa
Note that Eq.~(\ref{fullcon}) reduces to the valence contribution
Eq.~(\ref{valcon}) if one replaces $\Delta S_f(B)$ by unity.
Fig.~\ref{scatter} from Ref.~\cite{bruck} shows a scatter
plot of the condensate as a function of the change in the action $\Delta S_f(B)$ due to the magnetic field.
In this plot, 
the magnetic field strength is $eB\approx5$ GeV$^2$
and $T$ close to the transition temperature. 
Each point represents a gauge configuration and they were generated at vanishing magnetic field.
The plot suggests that larger values of the condensates
correspond to larger values of the weight 
$e^{-\Delta S_f(B)}$ and therefore suppresses the weight
of the associated gauge configuration. As a result, this
counteracts the valence effect, and leads to a decrease
in the critical temperature. For pion masses that are not
too large, it also
leads to a decrease of the condensate itself (see discussion
below).

\begin{figure}[htb]
\centering\includegraphics[width=0.45\textwidth]{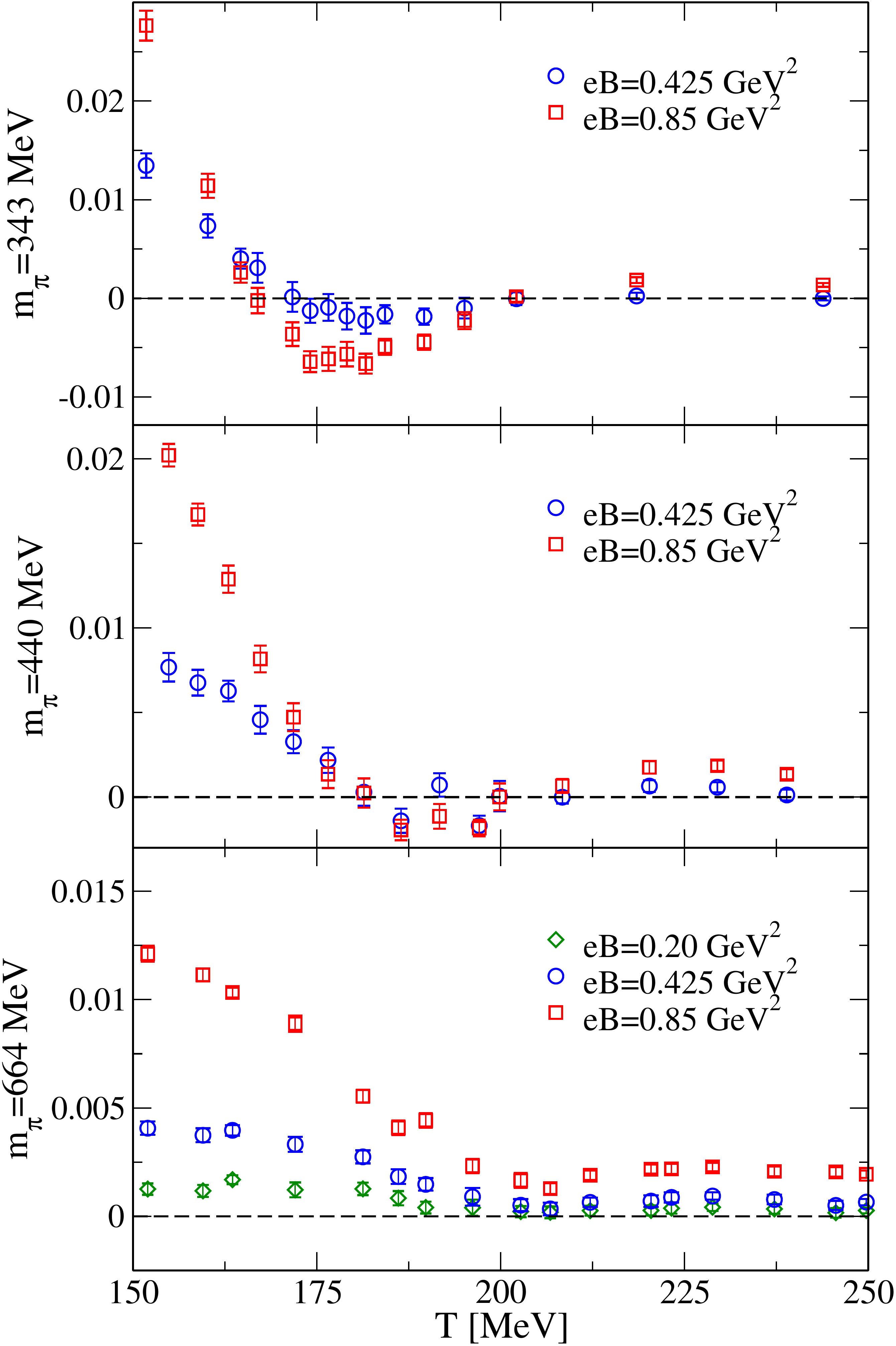}
\caption{Difference between the quark condensates at zero magnetic field and non-vanishing $B$ for three different values of the pion mass. See main text for details.
Figure taken from Ref.~\cite{delia2}.}
\label{b0}
\end{figure}

In Refs.~\cite{delia2,heavypi}, the effects of varying
the pion mass on the quark condensate as a function 
of the temperature have been studied in detail 
(again 2+1 flavors).
Fig.~\ref{b0} from~\cite{delia2}
shows the difference between the quark condensates 
as a function of the temperature
at $B=0$ and $eB=0.425$ GeV$^2$ (blue data points) and $eB=0.85$ GeV$^2$ (red points) for three values of the pion mass.
The authors find that the sea contribution is a decreasing 
function of $B$ around $T_c$ for the different values
of the pion masses, while the valence contribution is on the other hand an increasing function of the magnetic field 
for all temperatures and pion masses. 
In the upper panel it is clear that the 
sea contribution wins the competition around the transition
temperature implying inverse magnetic catalysis in the
strict sense of the word.
This effect can barely been seen in the middle panel and is
completely absent in the lower panel.
In other words, a decreasing function of the transition temperature does not imply that the chiral condensate decreases
as a function of temperature and it is therefore not clear
that the latter is the driving mechanism of the former~\cite{delia2}. This nontrivial behavior was
also demonstrated  in~\cite{heavypi}, where the
authors fixed the magnetic field to $eB=0.6$ GeV$^2$
and varied the pion mass. In QCD, there is 
inverse magnetic catalysis for pion masses up to 500 MeV, 
and magnetic catalysis for larger values~\footnote{
In Ref.~\cite{ding0}, the authors find no sign of inverse catalysis
in their $N_f=3$ simulations with a pion mass of $280$ MeV.}

We finally comment on the nature of the chiral transition
and the temperature as a function of $B$.
The simulations have been done with magnetic fields up to
$eB=1$ GeV$^2$. They all show an analytic crossover and
that the transition temperature is
a decreasing function of $B$. However, it has been conjectured
that the transition would start increasing again for
sufficiently large temperatures, a phenomenon dubbed 
delayed magnetic catalysis. In Ref.~\cite{ani}
the author went as high as $eB=3.25$ GeV in the simulations. 
The transition remains a crossover (albeit sharper), there is
no sign of delayed magnetic catalysis, and the chiral
and deconfinement transitions coincide.
The sharper crossover suggests that there may be a critical
point for even larger values of the magnetic field.
For asymptotically large fields, QCD can be mapped onto an
anistropic pure-glue theory~\cite{asymp}.
This theory was simulated on the lattice and strong evidence
for a first-order transition was found~\cite{ani}.
This implies the existence of a critical point and its
position was estimated to be at $eB\simeq 10$ GeV$^2$.

\section{Improvements of models}
\label{improve}
The failure of models to correctly describe the behavior of QCD around the critical
temperature, even after the introduction of the Polyakov loop, has lead to significant efforts to improve 
them, see e.g.~\cite{ferr2,fraga22,farias0,ayala,ferrer,jos,ferrer1,loewe1,maoinverse,skok,farias2}.  
The temperature $T_0$ that enters the
Polyakov-loop potential depends on the number of flavors, it is therefore
a reasonable assumption that it also depends on the magnetic field. In Ref.~\cite{ferr2}, the authors fitted the strange-quark susceptibilities
from their calculations in the entangled PNJL model to the lattice results of Ref.~\cite{bali0}. Their ansatz for the field dependence was a simple
polynomial in $(eB)^2$ up to quadratic order, giving two fitting parameters,
\bqa
T_0(eB)&=&T_0(0)+\zeta(eB)^2+\xi(eB)^4\;.
\eqa
An interesting feature here is that the model predicts a first-order
transition for magnetic fields larger than approximately 
$eB=0.25$ GeV$^2$. As mentioned above, such a critical point is expected in QCD, albeit at much larger magnetic fields~\cite{ani}.

\begin{figure}[htb]
\centering\includegraphics[width=0.5\textwidth]{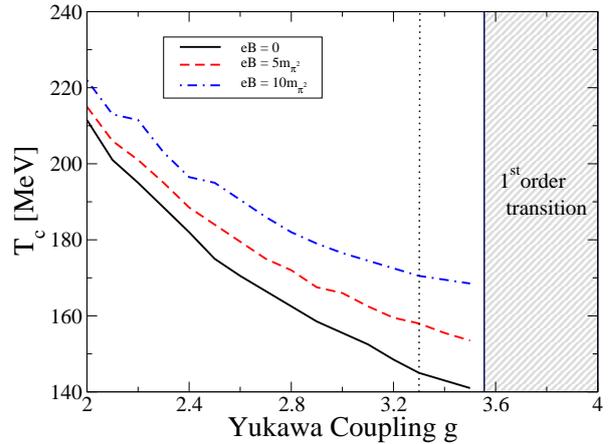}
\caption{$T_c$ as a function of the Yukawa coupling $g$ for various values of the magnetic field. See main text for details.
Figure taken from Ref.~\cite{fraga22}.}
\label{gragag}
\end{figure}

The crossover nature of the chiral transition was a guide
for the authors of Ref.~\cite{fraga22}, trying to incorporate
the decreasing behavior of the transition temperature in 
the (Polyakov-loop extended) two-flavor quark-meson model.
In their mean-field analysis, they allowed the Yukawa coupling
to vary with the magnetic field, $g=g(B)$ using the boundary
value $g(0)=3.3$. This value is indicated by the vertical dotted line 
in Fig.~\ref{gragag} and 
corresponds to a fixed quark mass in the vacuum.
The critical temperature as a function of the Yukawa coupling for three values of the magnetic  field is also shown in Fig.~\ref{gragag}. The grey shaded region indicates the values of $g$ for which the transition is first order.

However, to obtain a transition temperature which is 
decreasing with the magnetic field, any curve $g(B)$ must start at $g(0)=3.3$ and successively cross the dashed (red) and solid (black) curves. One therefore soon enters the shaded region which indicates a first-order transition in the QM model. Thus a function $g(B)$ can simply
not describe the correct $B$-dependence of 
the transition temperature, while at the same time
having a crossover transition.

Similar approaches have been used  in the NJL model allowing
the coupling $G$ to depend on both $T$ and $B$. 
For example, in Ref.~\cite{farias0}, the authors 
fix a set of parameters to get a reasonable fit to the
lattice data for the sum of the light quark condensates.
The form of the $B$-dependent coupling  was motivated
by the running of the coupling in QCD for strong magnetic 
fields~\cite{showrun}.

One particular appealing idea was recently put forward in Ref.~\cite{marko} (see also
Ref.~\cite{costi1}).
Only $T=0$ physics from the lattice is used as input to improve the model.
In other words, there is no fitting to lattice data for $T_c$ 
or a coupling that depends both upon $B$ and $T$.
The authors first performed a determination of the baryon spectrum at the physical point as a function of magnetic field using lattice simulations. The authors focused on strong magnetic fields, which are relevant for the phase diagram. Making the simple assumption that the
baryon masses can be written as the sum of the masses of their constituents, they derived $B$-dependent constituent quark masses. This was used as input in the PNJL model at zero temperature: Using the gap equation with the $B$-dependent quark masses, a $B$-dependent four-fermion coupling was obtained. The $B$-dependent constituent quark masses as well as $G(B)$
are decreasing functions of the magnetic field. For $B=0$, 
$M^2=0.097$ MeV$^2$ and $G(0)=12.8$ GeV$^{-2}$ and for the largest
magnetic field used ($|eB|=0.6$ GeV$^2$),
$M^2=0.079$ MeV$^2$ and $G(B)=6.7$ GeV$^{-2}$.

\begin{figure}[htb]
\centering\includegraphics[width=0.5\textwidth]{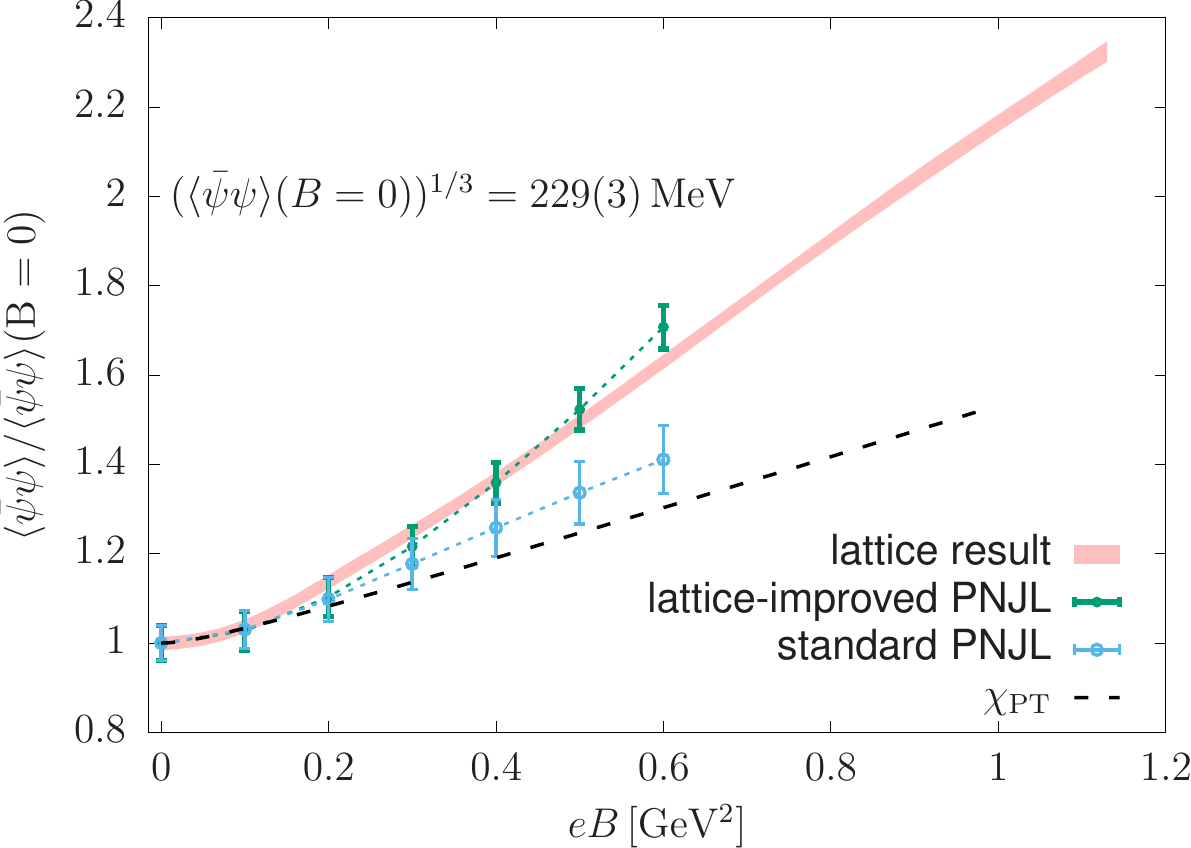}
\caption{Normalized average quark condensate in the PNJL model
as a function of $eB$. See main text for details.
Figure taken from Ref.~\cite{marko}.}
\label{b00b}
\end{figure}
Fig.~\ref{b00b} shows the normalized average quark condensate at $T=0$ as a function of the
magnetic field. The dashed black line is obtained from a next-to-leading order
(one-loop) calculation in chiral perturbation theory~\cite{smilga,werbos},
the light-blue points are obtained from the standard PNJL model, while
the green points are obtained from the lattice-improved PNJL model.
Finally, the red band shows the results from lattice simulations including errors.
The plot has several interesting features. Firstly, all low-energy 
approaches are in good agreement with lattice results for low values of the magnetic field.
For larger values, both $\chi$PT and PNJL underestimate the quark condensate.
This is in contrast with the lattice-improved PNJL model, which is in quantitative
very good agreement with the simulations. It would be of interest to see the 
predictions if one would include the effects of the $s$-quark
in the model calculations.

\begin{figure}[htb]
\centering\includegraphics[width=0.5\textwidth]{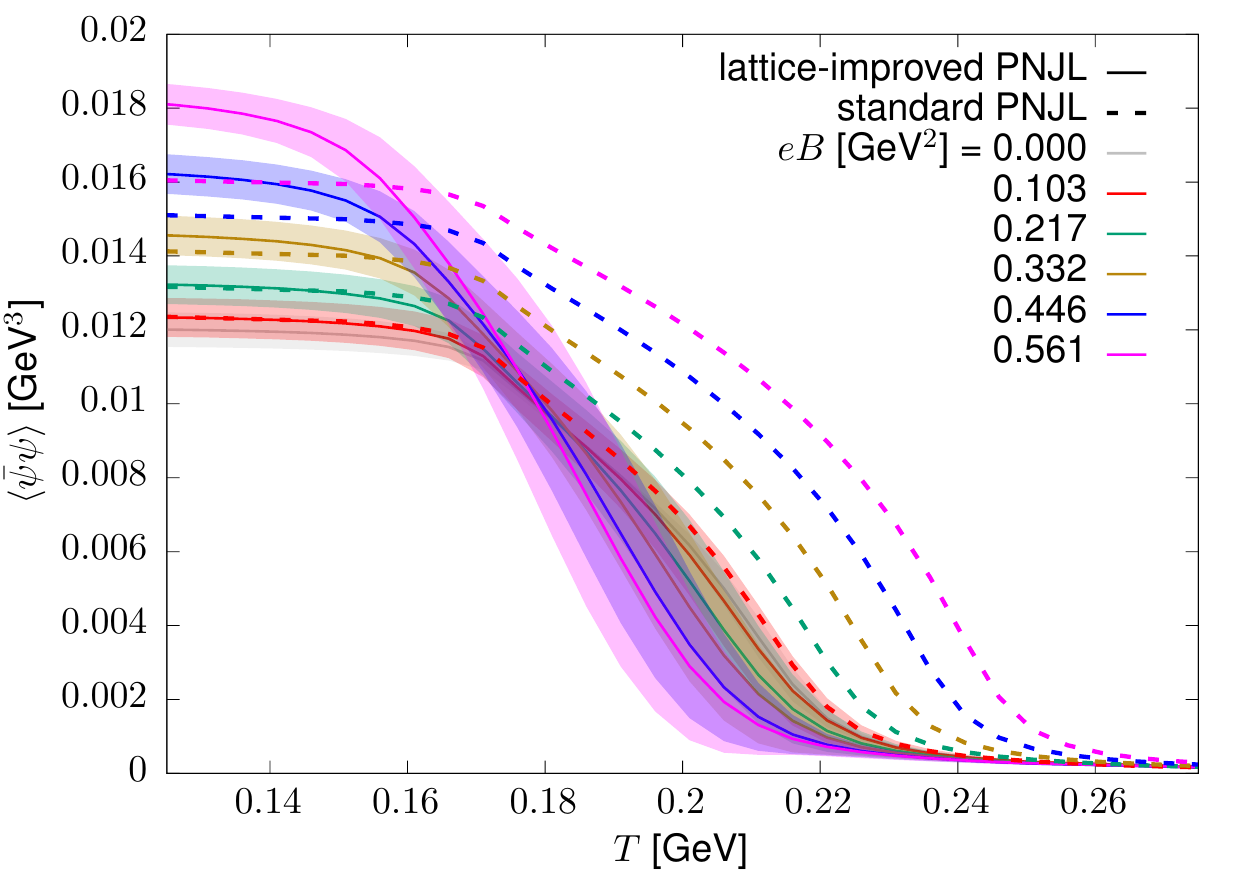}
\caption{Quark condensate in the PNJL model
as a function of $T$ for different
values of the magnetic field.
See main text for details.
Figure taken from Ref.~\cite{marko}.}
\label{b00}
\end{figure}

We next consider the finite-temperature calculations of Ref.~\cite{marko}.
Fig.~\ref{b00} shows the quark condensate as a function of $T$ for different strengths of the magnetic field. The dashed lines are the predictions from the
standard PNJL model without error bands, while the solid bands are the lattice-improved PNJL model predictions. We first note that quark condensate at $T=0$ for fixed
magnetic field is higher for the lattice-improved PNJL model in agreement with Fig.~\ref{b00b}.  
This behavior persists for low temperature, for larger temperatures, however, the condensate drops faster compared to that calculated using the standard PNJL model. Thus 
for fixed magnetic field, the transition temperature defined as the inflection point of the quark condensate moves to a lower value of $T$ compared to the PNJL model.
This in itself is not enough to conclude that we have
inverse magnetic catalysis, but the effect is
so strong that the inflection point moves to the
left as a function of $B$
so the transition temperature decreases.

\begin{figure}[htb]
\centering\includegraphics[width=0.45\textwidth]{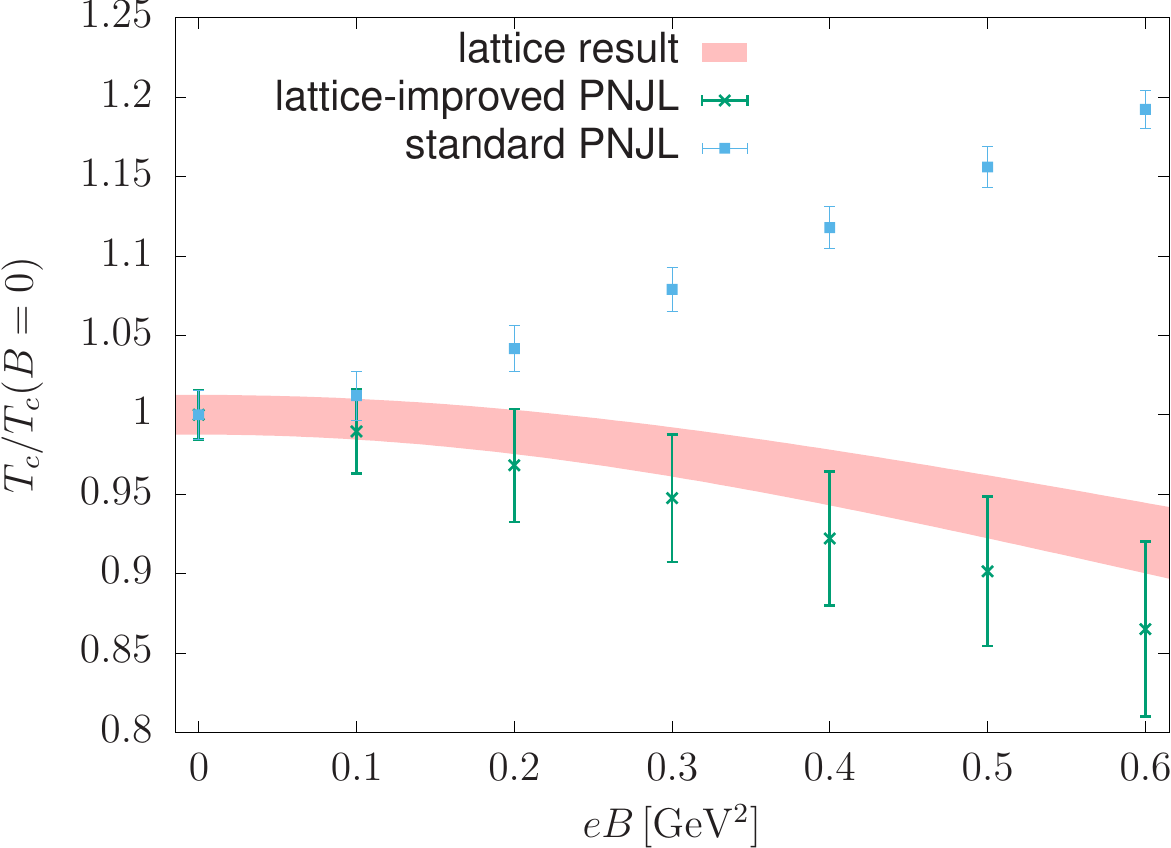}
\caption{Normalized transition temperature from the lattice and in the PNJL model as a function of $eB$.
See main text for details. Figure taken from Ref.~\cite{marko}.}
\label{b000}
\end{figure}

Fig.~\ref{b000} shows the normalized transition temperature for the chiral transition, as defined by the inflection point of the quark condensate, as a function of $eB$ in units
of GeV$^2$. The pink band is from the lattice results of Refs.~\cite{bali0}.
The width of the band indicates the errors of the simulations.
The light-blue points are the results of a calculation from the standard PNJL model showing that the transition temperature increases as the magnetic field grows.
This is in sharp contrast to the lattice-improved PNJL model, where the results are shown by the green points including uncertainties coming from the
lattice determination of the baryon masses. Given the large uncertainties, the
results for the transition temperature are in good agreement with the simulations.
Similarly, the analytic crossover found also agrees with lattice results.

\section{Summary and final remarks}
\label{summary}
The idea of magnetic catalysis at zero temperature
has been around for three decades
after its discovery in the NJL model. It is a 
robust phenomenon. For large values of the magnetic field, 
it can be understood in terms of dimensional reduction
from $3+1$ to $1+1$ dimensions
(or from 2+1 to 0+1 dimensions).
Lattice simulations in the last decade have 
improved our understanding of the effect significantly by
showing that both the valence and sea effect contribute
in a nontrivial way. 
Since the discovery of the
the decrease of the chiral transition temperature with increasing
magnetic field, a lot of work has been devoted to incorporate
this feature in models. This includes $T$ - and $B$-dependent couplings and  Polyakov-loop potentials.
Fitting parameters can be considered an indirect way of incorporating the sea effect and requires input from
lattice simulations at finite temperature.
In our opinion, a cleaner approach is provided by
the work~\cite{marko}, which uses lattice
input at $T=0$ only. In much the same way as one uses
experimentally measured meson masses in the vacuum, they use 
$B$-dependent baryon masses measured on the lattice as
input in their PNJL-model calculations,
although assuming that the baryon mass is the sum of its constituents perhaps is somewhat simplistic.

\section*{Acknowledgements}
The author would like to thank Prabal Adhikari, Patrick Kneschke, William Naylor, and Anders Tranberg for discussions and collaboration on related topics.
The author would like to thank Massimo D'elia, Gergely Endr\'{o}di, Eduardo Fraga,
and Vladimir Skokov for permission to use their figures.

\appendix
\section{Renormalization of the one-loop effective potential
in the quark-meson model} 
\label{appa}
In this appendix, we will discuss 
renormalization of the one-loop effective potential
in the quark-meson model
using the on-shell scheme.
The starting point is the one-loop contribution to the effective potential of a fermion of mass $m_f$ in a constant magnetic field, which is given by
\bqa
\nonumber
V_1&=&
-{|q_fB|\over2\pi}
\sum_{s=\pm1}\sum_{k=0}^{\infty}
\int_{p_{\parallel}}
\log\left[p_{\parallel}^2+m_f^2+|q_fB|(2k+1-s)\right]\;,
\\ &&
\label{deter}
\eqa
where $p_{\parallel}^2=p_0^2+p_z^2$ and the integral is
defined in $d$ dimensions using dimensional regularization,
\bqa
\int_p&=&
\left({e^{\gamma_E}\Lambda^2\over4\pi}\right)^{\epsilon}
\int{d^dp\over(2\pi)^{d}}\;.
\eqa
and where $\Lambda$ is the renormalization scale associated with the
$\overline{\rm MS}$-scheme.
The sum is over spin $s$ and Landau levels $k$.
The integral over $p_{\parallel}$ can be evaluated using dimensional regularization
in $d=2-2\epsilon$ dimensions. 
The result is
\bqa
V_1&=&
{2|q_fB|\over(4\pi)^2}\Gamma(-1+\epsilon)
\left({e^{\gamma_E}\Lambda^2}\right)^{\epsilon}
\sum_{s=\pm1}\sum_{k=0}^{\infty}M_B^{2-2\epsilon}\;,
\eqa
where $M_B^2=m_f^2+|q_fB|(2k+1-s)$.
The sum over spin $s$ and Landau levels $n$ can be expressed in terms
of the Hurwitz $\zeta$-function as
\bqa\nonumber
\sum_{s=\pm1}\sum_{k=0}^{\infty}M_B^{2-2\epsilon}&=&
2(2|q_fB|)^{1-\epsilon}
\sum_{k=0}^{\infty}\left[k+{m_f^2\over2|q_fB|}\right]^{1-\epsilon}
-m_f^{2-2\epsilon}
\\
&=&2(2|q_fB|)^{1-\epsilon}\zeta(-1+\epsilon,x_f)-m_f^{2-2\epsilon}\;,
\eqa
where $x_f={m_f^2\over2|q_fB|}$ is a dimensionless variable.
The effective potential  can then be written as
\bqa\nonumber
V_1&=&{8(q_fB)^2\over(4\pi)^2}
\left({e^{\gamma_E}\Lambda^2\over2|q_fB|}\right)^{\epsilon}
\Gamma(-1+\epsilon)\bigg[\zeta(-1+\epsilon,x_f)
\\ &&
  -{1\over2}x_f^{1-\epsilon}\bigg]\;.
  \label{f1}
\eqa
We next expand the result (\ref{f1}) in powers of $\epsilon$ to order $\epsilon^0$. This yields
\bqa\nonumber
{V}_1&=&
{1\over(4\pi)^2}\left({\Lambda^2\over2|q_fB|}\right)^{\epsilon}
\bigg[\left({2(q_fB)^2\over3}+m_f^4\right)\left({1\over\epsilon}+1\right)
 \\ && 
-8(q_fB)^2\zeta^{(1,0)}(-1,x_f)-2|q_fB|m_f^2\log x_f
\bigg]\;.
\label{div}
\eqa
For renormalization purposes, 
it is convenient to isolate in Eq.~(\ref{div}) the terms in the functional determinant
that equal the $B=0$ result, cf. Eq.~(\ref{logb0}).
Adding the tree-level potential $V_0$, setting $m_f=\Delta$,
and summing over quark flavors and colors yields
\bqa\nonumber
V_{0+1}&=&
{1\over2}B^2+{1\over2}{m^2\over g^2}\Delta^2
+{\lambda\over24g^4}\Delta^4-{h\over g}\Delta
\\ &&
\nonumber
+{2N_c\Delta^4\over(4\pi)^2}\left[{1\over\epsilon}+{3\over2}
+\log{\Lambda^2\over\Delta^2}\right]
\\ && \nonumber
+{N_c\over(4\pi)^2}\sum_f{2(q_fB)^2\over3}\left[{1\over\epsilon}
+\log{\Lambda^2\over2|q_fB|}
\right]
\\ &&
\nonumber
-{8N_c\over(4\pi)^2}\sum_f(q_fB)^2\left[
{\zeta^{(1,0)}(-1,x_f)}
+{1\over4}x_f^2
\right. \\ &&\left.
-{1\over2}x_f^2\log x_f+{1\over2}x_f\log x_f
-{1\over12}
\right]\;.
\label{div2}
\eqa
Eq.~(\ref{div2}) has simple poles in epsilon. The pole
proportional to $(q_fB)^2$
is eliminated by wavefunction renormalization of $B$ 
and the charge $q_f$ such that $q_fB$ is invariant.
The pole propertional to $\Delta^4$ is eliminated by renormalization of the parameters in the Lagrangian.
The bare parameters
are replaced by the parameters in $\overline{\rm MS}$-scheme, e.g. $m^2\rightarrow m^2_{\ms}+\delta m^2_{\ms}$ and the running parameters given 
by Eqs.~(\ref{sol1})--(\ref{sol5}) are substituted
into the renormalized expression for the effective potential. The couplings at the reference scale
$\Lambda_0$ are determined 
using Eqs.~(\ref{osm1})--(\ref{osm4}) and expressed
in terms of physical quantities.
The result is Eq.~(\ref{fullb}).

\setcounter{equation}{0}
\section{Parameter fixing}
\label{appb}
The mass of a particle is given by the pole of the propagator.
In the on-shell scheme, the sum of the self-energy
evaluated on-shell and the counterterms
vanishes,
\bqa
\Sigma_{\sigma,\pi}^{\rm }(p^2=m^2_{\sigma,\pi})
+{\rm counterterms}
&=&0\;.
\eqa
In addition, the residue of the propagator evaluated on shell is unity. This implies
\bqa
\Sigma^{\prime}(p^2=m_{\sigma,\pi}^2)+{\rm counterterms}&=&0\;.
\eqa
The self-energies are
\bqa
\Sigma_{\sigma}(p^2)&=&-8ig^2N_c\left[
A(m_q^2)-\mbox{$1\over2$}\left(p^2-4m_q^2\right)B(p^2)\right]
\;,
\label{selfie1}
\\ 
\Sigma_{\pi}(p^2)&=&-8ig^2N_c\left[
A(m_q^2)-\mbox{$1\over2$}p^2B(p^2)\right]
\;,
\label{selfie2}
\eqa
where the integrals $A(m^2)$ and $B(p^2)$ are defined in Eqs.~(\ref{adef})--~(\ref{bdef}).
We have omitted the
tadpole diagram which is one-particle reducible. It is cancelled by a counterterm, when we impose the condition that $\phi_0=f_{\pi}$. The counterterms are given by the expressions
\bqa
\delta m_{\sigma,\pi}^2&=&
-\Sigma_{\sigma}(p^2)
\Big|_{p^2=m_{\sigma,\pi}^2}\;,
\hspace{1mm}
\delta Z_{\sigma,\pi} =
\Sigma_{\sigma,\pi}^{\prime}(p^2)\Big|_{p^2=m_{\sigma,\pi}^2}\;. 
\;.
\eqa
This yields
\bqa
\delta m_{\sigma}^2&=&
8ig^2N_c\left[A(m_q^2)-\mbox{$1\over2$}(m_{\sigma}^2-4m_q^2)B(m_{\sigma}^2)
\right]\;,
\\ 
\delta m_{\pi}^2&=&8ig^2N_c
\left[
A(m_q^2)-\mbox{$1\over2$}m_{\pi}^2B(m_{\pi}^2)\right]\;,
\\
\delta Z_{\sigma}&=&
4ig^2N_c\left[B(m_{\sigma}^2)+(m_{\sigma}^2-4m_q^2)B^{\prime}(m_{\sigma}^2)
\right]\;,
\\
\delta Z_{\pi}&=&4ig^2N_c\left[B(m_{\pi}^2)
+m_{\pi}^2B^{\prime}(m_{\pi}^2)\right]
\label{zpi}
\;,\\
\delta t&=&-8ig^2N_cf_{\pi}A(m_q^2)\;,
\eqa
where we have added the counterterm $\delta t$ for the one-point function. 
We next need to relate the above counterterms to the counterterms of the parameters of the Lagrangian. These relations follow immediately from
Eqs.~(\ref{tr1})--(\ref{tr4}),
\bqa
\label{rela1}
\delta m^2_{\os}&=&-{1\over2}(\delta m_{\sigma}^2-3\delta m_{\pi}^2)\;,\\
\delta\lambda_{\os}&=&3{(\delta m_{\sigma}^2-\delta m_{\pi}^2)\over f_{\pi}^2}
-\lambda{\delta f_{\pi}^2\over f_{\pi}^2}\;,\\
\delta g^2_{\os}&=&{\delta m_q^2\over f_{\pi}^2}
-g^2{\delta f_{\pi}^2\over f_{\pi}^2}\;.
\label{rela2}
\eqa
In the large-$N_c$ limit, $\delta m_q^2=0$ implying that 
$\delta g^2_{\os}=-g^2{\delta f_{\pi}^2\over f_{\pi}^2}$. There is also no loop correction to the quark-pion vertex
and $\delta Z_{\psi}=1$. This implies
that the associated counterterms must cancel as well, leading to
$\delta g^2_{\os}=-g^2\delta Z_{\pi}^{\os}$. 
We can therefore write
\bqa
\label{rela4}
\delta\lambda_{\os}&=&3{(\delta m_{\sigma}^2-\delta m_{\pi}^2)\over f_{\pi}^2}-
\lambda\delta Z_{\pi}^{\os}\;.
\eqa
The counterterm $\delta h_{\os}$ is found from the one-point function. At tree level, we have $t=h-m_{\pi}^2f_{\pi}=0$, which yields
$\delta t=\delta h_{\os}-\delta m_{\pi}^2f_{\pi}-m_{\pi}^2\delta f_{\pi}=$ or $\delta h_{\os}=\delta t+\delta m_{\pi}^2f_{\pi}+{1\over2}m_{\pi}^2f_{\pi}\delta Z_{\pi}^{\os}$.
Finally, we need the counterterm for the electromagnetic field,
\bqa
\delta Z_A^{\os}&=&i{N_c\over(4\pi)^2}\sum_f{4q_f^2\over3}B(0)\;,
\label{b22}
\eqa
with the integral $B(p^2)$ defined in Eq.~(\ref{bdef}).
Renormalization is then carried out by making the substitution
$B^2\rightarrow B^2(1+\delta Z_A^{\os})$.
We note that the on-shell scheme is not well-defined
when the fermions are massless. In that case, the (modified) minimal subtraction scheme may be used.
Since the bare parameters are independent
of the renormalization scheme, we can immediately
write down the relations between
the renormalized parameters in the on-shell and $\overline{\rm MS}$ schemes. 
For example $g_{\ms}^2+\delta g_{\ms}^2=g^2_{\os}+\delta g^2_{\os}$. From Eq.~(\ref{zpi}), we find
\bqa\nonumber
\delta g^2_{\rm os}&=&{4g^4N_c\over(4\pi)^2}\left[
{1\over\epsilon}+\log{\Lambda^2\over m_q^2}+F(m_{\pi}^2)
+m_{\pi}^2F^{\prime}(m_{\pi}^2)
\right]\;,
\\ &&
\eqa
where
$F(p^2)$ and $F^{\prime}(p^2)$ are defined in Eqs.~(\ref{fdef})--(\ref{fpdef}).
The counterterm in the $\overline{\rm MS}$-scheme is simply the
pole part,
$\delta g^2_{\ms}={4N_cg^4\over(4\pi)^2}{1\over\epsilon}$.
From this, one finds the running coupling $g^2_{\ms}$ using
$g_{\ms}^2=g^2_{\os}+\delta g^2_{\os}-\delta g_{\ms}^2$ and
given by Eq.~(\ref{osm3}). The running parameters are
\begin{widetext}
\bqa 
\label{osm1}
m_{\ms}^2&=&-{1\over2}\left(m_{\sigma}^2-3m_{\pi}^2\right)
-{2m_q^2N_c\over(4\pi)^2f_{\pi}^2}\left[
  \left(m_{\sigma}^2-3m_{\pi}^2\right)\log{\Lambda^2\over m_q^2}
  +4m_q^2+\left(m_{\sigma}^2-4m_q^2\right)F(m_{\sigma}^2)
  -3m_{\pi}^2F(m_{\pi}^2)
  \right]\;,\\ \nonumber
\lambda_{\ms}&=&{3\left(m_{\sigma}^2-m_{\pi}^2\right)\over f_{\pi}^2}
+{12g^2N_c\over(4\pi)^2f_{\pi}^2}\bigg[
2\left(m_{\sigma}^2-m_{\pi}^2-2m_q^2\right)\log{\Lambda^2\over m_q^2}
+\left(m_{\sigma}^2-4m_q^2\right)F(m_{\sigma}^2)
\\ &&
+\left(m_{\sigma}^2-2m_{\pi}^2\right)F(m_{\pi}^2)
+\left(m_{\sigma}^2-m_{\pi}^2\right)m_{\pi}^2F^{\prime}(m_{\pi}^2)
\bigg]\;,  \\
\label{osm3}
g_{\ms}^2&=&{m_q^2\over f_{\pi}^2}\left\{1
+{4m_q^2N_c\over(4\pi)^2f_{\pi}^2}\left[
  \log{\Lambda^2\over m_q^2}
  +F(m_{\pi}^2)+m_{\pi}^2F^{\prime}(m_{\pi}^2)\right]\right\}\;,\\
\label{osm4}
h_{\ms}&=&m_{\pi}^2f_{\pi}\left\{1
+{2m_q^2N_c\over(4\pi)^2f_{\pi}^2}\left[
  \log{\Lambda^2\over m_q^2}
  +F(m_{\pi}^2)-m_{\pi}^2F^{\prime}(m_{\pi}^2)\right]\right\}\;.
\label{osm5}
\eqa

The running parameters satisfy renormalization group equations
that follow from Eqs.~(\ref{osm1})--(\ref{osm4}) upon
differentiation with respect to $\Lambda$. The solutions are
\end{widetext}
\bqa
\label{sol1}
m_{\ms}^2(\Lambda)&=&{m_0^2\over1-{4g_0^2N_c\over(4\pi)^2}
\log{\Lambda^2\over m_q^2}}\;,
\\
g_{\ms}^2(\Lambda)&=&
{g_0^2\over1-{4g_0^2N_c\over(4\pi)^2}
\log{\Lambda^2\over m_q^2}
}\;,
\\
\label{sol4}
\lambda_{\ms}(\Lambda)&=&{\lambda_0-{48g_0^4N_c\over(4\pi)^2}
\log{\Lambda^2\over m_q^2}
\over\left(1-{4g_0^2N_c\over(4\pi)^2}
\log{\Lambda^2\over m_q^2}
\right)^2}\;, \\
h_{\ms}(\Lambda)&=&
{h_0\over1-{2g_0^2N_c\over(4\pi)^2}
\log{\Lambda^2\over m_q^2}
}\;,
\label{sol5}
\eqa
where $m_0^2$, $g_0^2$, $\lambda_0$, and $h_0$
are the values of the running mass and couplings at the scale 
$\Lambda_0$ determined by
\bqa
\left[
  \log{\Lambda_0^2\over m_q^2}
  +F(m_{\pi}^2)+m_{\pi}^2F^{\prime}(m_{\pi}^2)\right]&=&0\;.
\eqa
This equation in conjunction with
Eqs.~(\ref{osm1})--(\ref{osm4})
can be used to determine the values of the couplings at the scale $\Lambda_0$ expressed in terms of
physical quantities. For example, it follows that
$g^2_0=g^2_{\ms}(\Lambda_0)={m_q^2\over f_{\pi}^2}$.

We need a few divergent integrals 
space in four dimensions. 
Going to Euclidean space via Wick rotation, we can use dimensional regularization
in $d=4-2\epsilon$ dimensions. 
The integrals needed are
\bqa
\label{logb0}
\int_k\log\left[k^2+m^2\right]&=&
-{m^4\over2(4\pi)^2}\left({\Lambda^2\over m^2}\right)^{\epsilon}
\left[{1\over\epsilon}+{3\over2}+{\cal O}(\epsilon)\right]\;,\\
\nonumber
A(m^2)&=&\int_k{1\over k^2-m^2}
\\ &=&
{im^2\over(4\pi)^2}\left({\Lambda^2\over m^2}\right)^{\epsilon}
\left[{1\over\epsilon}+1+{\cal O}(\epsilon)\right]\;,
\label{adef}
\\ \nonumber
B(p^2)&=&\int_k{1\over(k^2-m_q^2)[(k+p)^2-m_q^2)}
\\ \nonumber
&=&{i\over(4\pi)^2}\left({\Lambda^2\over m_q^2}\right)^{\epsilon}
\left[{1\over\epsilon}+F(p^2)+{\cal O}(\epsilon)\right]\;,
\\ &&
\label{bdef}
\eqa
where $\Lambda$ is the renormalization scale associated with the $\overline{\rm MS}$ scheme
and 
\bqa
F(p^2)&=&2-2r\arctan\left({1\over r}\right)\;,
\label{fdef}
\\
B^{\prime}(p^2)&=&
F^{\prime}(p^2)={4m_q^2\over p^4r}\arctan\left({1\over r}\right)
-{1\over p^2}\;,
\label{fpdef}
\eqa
where $r=\sqrt{{4m_q^2\over p^2}-1}$.



\begin{thebibliography}{}
\bibitem{colorfirst}
D. Bailin and A. Love,
Phys. Rept. {\bf 107}, 325 (1984).

\bibitem{bcs}
J. Bardeen, L. N. Cooper and J. R. Schrieffer, Phys. Rev.
\textbf{106}, 162 (1957).


\bibitem{raja}
K. Rajagopal and F. Wilczek, 
At the frontier of particle physics, Vol. 3
(World Scientific, Singapore, p 2061) (2001).
\bibitem{alford}
M. G. Alford, A. Schmitt, K. Rajagopal, and T. Sch\"afer,
Rev. Mod. Phys. {\bf 80}, 1455 (2008).

\bibitem{fukurev}
K. Fukushima and  T. Hatsuda,
Rept. Prog. Phys. {\bf 74}, 014001 (2011).

\bibitem{rob1}
L. McLerran and R. D. Pisarski, Nucl. Phys. A {\bf 796} 83 (2007).

\bibitem{kar1}
D. E. Kharzeev, L. D. McLerran, and H. J. Warringa,
Nucl. Phys. A {\bf 803}, 227 (2008).

\bibitem{duncan}
R. C. Duncan and C. Thompson, Astrophys. J. {\bf 392}, L9 (1992).




\bibitem{review}
D. Kharzeev, K. Landsteiner, A. Schmitt, and H.-U. Yee,
Lect. Notes Phys. {\bf 871}, 1 (2013).



\bibitem{dima}
D. E. Kharzeev, Ann. Rev. Nucl. Part. Sci. {\bf 65}, 193 (2015).

\bibitem{ourrev}
J. O. Andersen, W. R. Naylor, and A. Tranberg,
Rev. Mod. Phys. {\bf 88}, 025001 (2016).



\bibitem{fariasrev}
A. Bandyopadhyay and R. L. S. Farias, e-print: 2003.11054 [hep-ph].

\bibitem{lemmer}
S. P. Klevansky and R. H. Lemmer,
Phys. Rev. D 39, 3478 (1989).

\bibitem{sugu}
H. Suganuma and T. Tatsumi, Annals Phys. {\bf 208}, 470 (1991).

\bibitem{klim1}
K. G. Klimenko, Z. Phys. C {\bf 54}, 323 (1992).
\bibitem{klim2}
K. G. Klimenko,
Theor. Math. Phys. {\bf 89}, 1161 (1992).
\bibitem{klim3}
K. G. Klimenko, Theor. Math. Phys. {\bf 90}, 3 (1992).

\bibitem{qedgus}
V. P. Gusynin, V. A. Miransky, and I. A. Shovkovy, 
Phys. Rev. Lett. {\bf 73}, 3499 (1994).
\bibitem{smilga}
I. A. Shushpanov and A. V. Smilga, Phys. Lett. B {\bf 402}, 351 (1997).

\bibitem{werbos}
T. D. Cohen, D. A. McGady, and E. S. Werbos, Phys. Rev. C {\bf 76}, 055201, (2007).

\bibitem{werbos2}
E. Werbos, Phys. Rev. C {\bf 77}, 065202 (2008).

\bibitem{andreas}
A. Haber, F. Preis, and A. Schmitt,
Phys. Rev. D {\bf 90}, 125036 (2014).

\bibitem{cherno}
P. Buividovich, M. Chernodub, E. Luschevskaya, and M. Polikarpov,
Phys. Lett. B {\bf 682}, 484 (2010).

\bibitem{chern2}
P. Buividovich, M. Chernodub, E. Luschevskaya, and M. Polikarpov,
Nucl. Phys. B {\bf 826}, 313 (2010).

\bibitem{braguta}
V. V. Braguta, P. V. Buividovich, T. Kalaydzhyan, S. V. Kuznetsov,  and M. I. Polikarpov, Phys. Atom. Nucl. {\bf 75}, 488 (2012).

\bibitem{delia}
M. D’Elia and F. Negro,
Phys. Rev. D {\bf 83} 114028 (2011).

\bibitem{quarkcon}
G. S. Bali, F. Bruckmann, G. Endrodi, Z. Fodor, S. D. Katz, and A. Schäfer, Phys. Rev. D {\bf 86}, 071502(R) (2012).

\bibitem{sirlin}
  A. Sirlin, Phys. Rev. D {\bf 22}, 971 (1980).
\bibitem{sirlin2}
  A. Sirlin, Phys. Rev. D {\bf 29}, 89 (1984).
\bibitem{bohm}
  M. Bohm, H. Spiesberger, and W. Hollik, Fortsch. Phys. {\bf 34}, 687 (1986).
  
\bibitem{hollik}
W Hollik, Fortsch. Phys. {\bf 38}, 165 (1990).

\bibitem{allofus}
P. Adhikari, J. O. Andersen, and P. Kneschke,
Phys. Rev. D {\bf 98}, 074016 (2018).

\bibitem{reg1}
D.  Ebert,  K.  G.  Klimenko,  M.  A.  Vdovichenko,  and 
A.  S.  Vshivtsev,   
Phys.  Rev.  D {\bf 61},   025005  (2000).


\bibitem{menes} 
D. P. Menezes, M. Benghi Pinto, S. S. Avancini, A. Pérez Martínez, and C. Providência,
Phys. Rev. C {\bf 79}, 035807 (2009).

\bibitem{farias} 
R. L. S. Farias, K. P. Gomes, G. Krein, and M. B. Pinto,
Phys. Rev. C {\bf 90}, 025203 (2014).
 



\bibitem{gus2}
V. P. Gusynin, V. A. Miransky, and I. A. Shovkovy,
Nucl. Phys. B {\bf 462}, 249 (1996).   

\bibitem{coleman}
S. Coleman, Commun. Math. Phys. {\bf 31}, 259 (1973).

\bibitem{swing}
J. Schwinger,  Phys. Rev.82, 664 (1951).
 
\bibitem{farreg}
S. S. Avancini, R. L.S. Farias,  N. N. Scoccola, and W. R. Tavares,
Phys. Rev. D {\bf 99}, 116002 (2019).
\bibitem{farreg0}
S. S. Avancini, R. L. S. Farias, and W. R. Tavares,
Phys. Rev. D {\bf 99}, 056009 (2019).

\bibitem{hrg}
G.  Endr\'{o}di,
JHEP {\bf 04}, 023 (2013).

\bibitem{para}
G. S. Bali, F. Bruckmann, Endr\'{o}di, F. Gruber, and A. Schäfer, 
JHEP {\bf 04}, 130 (2013).



\bibitem{sidney}
S. S. Avancini, R. L. S. Farias, M. B. Pinto, T. E. Restrepo, and W. R. Tavares,
Phys. Rev. D {\bf 103}, 056009 (2021).


\bibitem{banks}
T. Banks and A. Casher, 
Nucl. Phys. B {169}, 103 (1980).



\bibitem{bruck}
F. Bruckmann, G.  Endr\'{o}di, and T. G. Kovacs,    
JHEP {\bf 04}, 112 (2013).


\bibitem{fraga0}
E. S. Fraga and A. J Mizher,
Phys. Rev. D {\bf 78}, 025016 (2008).



\bibitem{fraga}
A. J. Mizher, M. N. Chernodub, and E. S. Fraga
Phys. Rev. D {\bf 82}, 105016 (2010).

\bibitem{gatto}
R. Gatto and M. Ruggieri,
Phys. Rev. D {\bf 82}, 054027 (2010).



\bibitem{gatto2}
R. Gatto and M. Ruggieri, Phys. Rev. D {\bf 83}, 034016 (2011).


\bibitem{skokov}
V. V. Skokov,
Phys. Rev. D {\bf 85},  034026 (2012).
\bibitem{marco1}
M. Ruggieri, M. Tachibana, and V. Greco,
JHEP {\bf 07}, 165 (2013).


\bibitem{william}
J. O. Andersen, W. R. Naylor, and  A. Tranberg,
JHEP {\bf 04}, 187 (2014).




\bibitem{aoki}
Y. Aoki, Z. Fodor, S. Katz, and K. Szabo, 
Phys. Lett. B {\bf 643}, 46 (2006).

\bibitem{aoki2}
Y. Aoki, S. Borsanyi, S. Durr, Z. Fodor, S.D. Katz et al ,
JHEP 0906, 088 (2009).

\bibitem{borsa}
S.  Borsanyi et  al (Wuppertal-Budapest  Collaboration),
JHEP 1009, 073 (2010).

\bibitem{baza0}
A.  Bazavov,  T.  Bhattacharya,  M.  Cheng,  C.  DeTar,
H. Ding et al, Phys. Rev. D {\bf 85}, 054503 (2012).

\bibitem{baza}
A. Bazavov et al, Phys. Rev. D {\bf 93}, 114502 (2016).

\bibitem{yaffe}
  L. G. Yaffe and B. Svetitsky,
  Phys. Rev. D {\bf 26}, 963 (1982), Nucl. Phys. B {\bf 210}, 423 (1982).

\bibitem{polyakov1}
K. Fukushima,
Phys. Lett. B {\bf 591}, 277 (2004).

\bibitem{glue}
Karsch, F., E. Laermann,  and A. Peikert,
Nucl. Phys. B {\bf 605}, 579 (2001).


\bibitem{ratti}
C. Ratti, M. A. Thaler, and W. Weise, 
Phys. Rev. D {\bf 73}, 014019 (2006).

\bibitem{sjafer}
B. J. Schaefer, J. M. Pawlowski, and J. Wambach, 
Phys. Rev. D {\bf 76}, 074023 (2007).

\bibitem{ratti2}
C. Ratti, S. Roessner, M. A. Thaler, and W. Weise, 
Eur. Phys. C {\bf 49}, 213 (2007).

\bibitem{ratti3}
C. Ratti, S. Roessner, and W. Weise,
Phys.Rev. D {\bf 75}, 034007 (2007).

\bibitem{wetterich}
  C. Wetterich, Phys. Lett. B {\bf 301}, 90 (1993).
  
\bibitem{kamikado}
K. Kamikado and T. Kanazawa,
JHEP {\bf 03}, 009 (2014).
  
\bibitem{delia1}
M. D’Elia, S.  Mukherjee, and F. Sanfilippo
Phys. Rev. D {\bf 82}, 051501(R) (2010).

\bibitem{bali0}
G. S. Bali, F. Bruckmann, G. Endr\'{o}di, Z. Fodor, S. D. Katz, S. Krieg et al,
JHEP {\bf 02}, 044 (2012).

\bibitem{ani}
G. Endrodi, JHEP 15 {\bf 07}, 173 (2015).
\bibitem{delia2}
M. D'Elia, F. Manigrasso, F. Negro, and F. Sanfilippo,
Phys. Rev. D {\bf 98}, 054509 (2018).

\bibitem{heavypi}
G. Endrodi, M. Giordano, S. D. Katz, T. G. Kovács,
and F. Pittler,
JHEP {\bf 07}, 007 (2019).

\bibitem{ding0}
H.-T. Ding, C. Schmidt, A. Tomiya, and X.-D. Wang
Phys. Rev. D {\bf 102}, 054505 (2020).

\bibitem{ding1}
H.-T.  Ding,  S.-T.  Li,  A.  Tomiya,  X.-D.  Wang, 
and  Y.  Zhang, 
Phys. Rev. D {\bf 102},  054505  (2020).






\bibitem{asymp} 
V. P. Gusynin, V. A. Miransky, and I. A. Shovkovy,
Nucl. Phys. B {\bf 462}, 249 (1996).


    

\bibitem{ferr2}
M. Ferreira, P. Costa, D. P. Menezes, C. Providência, and N. N. Scoccola, Phys. Rev. D {\bf 89}, 016002 (2014).



 

\bibitem{fraga22}
E. S. Fraga, B. W. Mintz, and J. Schaffner-Bielich, 
Phys. Lett. {\bf B} 731, 154 (2014).


\bibitem{farias0}
R. L. S. Farias, K. P. Gomes, G. Krein, and M. B. Pinto
Phys. Rev. C {\bf 90}, 025203 (2014).


\bibitem{ayala} 
A. Ayala, M. Loewe, A. J. Mizher, and R. Zamora
Phys. Rev. D 90, 036001  (2014).

\bibitem{ferrer}
E. J. Ferrer, V. de la Incera, and X. J. Wen,
Phys. Rev. D {\bf 91}, 054006 (2015)

\bibitem{jos}
L. Yu, J. Van Doorsselaere, and M. Huang
Phys. Rev. D {\bf 91}, 074011 (2015).

\bibitem{ferrer1}
E. J. Ferrer, V. de la Incera, and X. J. Wen,
Phys. Rev. D {\bf 91}, 054006  (2015).

\bibitem{loewe1}
A. Ayala, C. A. Dominguez, L. A. Hernández, M. Loewe, and R. Zamora,
Phys. Lett. B {\bf 759} , 99 (2016).


\bibitem{maoinverse}
S. Mao and X. Jiaotong,
Phys. Lett. B {\bf 758}, 195 (2016).

\bibitem{skok}
V. P. Pagura, D. Gómez Dumm, S. Noguera, and N. N. Scoccola,
Phys. Rev. D {\bf 95}, 034013 (2017).

\bibitem{farias2}
R. L. S. Farias, V. S. Timóteo, S. S. Avancini, M. B. Pinto, and G. Krein, EPJA {\bf 53}, 101 (2017).

\bibitem{showrun} 
V. A. Miransky and I. A. Shovkovy, Phys. Rev. D {\bf 66},
045006 (2002).

\bibitem{marko}
G.  Endr\'{o}di and G. Marko,
JHEP {\bf 08}, 036 (2019).

\bibitem{costi1} 
J. Moreira, P. Costa, and T. E. Restrepo,
Phys. Rev. D {\bf 102}, 014032 (2020) and 
Eur. Phys. J. A {\bf 57}, 4 123 (2021).


 

 
 \end{thebibliography}
\end{document}